\newif\ifMNRAS
\MNRASfalse 

\PassOptionsToPackage{pdfpagelabels=false}{hyperref} 

\ifMNRAS
    \documentclass[fleqn,usenatbib,useAMS]{mnras}
    \usepackage{graphicx}
\else
       
\documentclass[nolinenumber]{aastex63}    
\fi

\usepackage{acronym}
\usepackage{hyperref}
\usepackage{amsmath,amsfonts,amssymb}
\usepackage{mathrsfs}
\usepackage{mathtools}
\usepackage[utf8]{inputenc}

\usepackage[normalem]{ulem}
\newcommand{\adam}[1]{{#1}}

\newcommand{\equ}[1]{
\begin{equation}
\begin{split}
#1
\end{split}
\end{equation}
}

\date{Received December 8, 2025}

\shorttitle{Fast-Flavor Conversion in CCSNe}

\begin{document}
\title{The Effect of the Fast-Flavor Instability on Core-Collapse Supernova Models: II. Quasi-Equipartition and the Impact of Various Angular Reconstruction Methods} 
\correspondingauthor{Tianshu Wang}
\email{tianshuw@berkeley.edu}
\author[0000-0002-0042-9873]{Tianshu Wang}
\affiliation{Department of Physics, University of California, Berkeley, CA, 94720-7300 USA}
\author[0000-0002-3099-5024]{Adam Burrows}
\affiliation{Department of Astrophysical Sciences, Princeton University, NJ 08544, USA}
   
\begin{abstract}
In this work, we explore in a consistent fashion the effects of fast flavor conversion (FFC) in 1D and 2D core-collapse supernova (CCSN) simulations. In addition, we investigate the impact of various angular reconstruction methods and compare the ``3-species'' and ``4-species'' neutrino transport schemes. We find that the FFC effects are insensitive to the different methods tested and that the FFC alters supernova hydrodynamics is only minor ways. We also present a ``quasi-equipartition'' approximation which can be used to estimate the FFC-altered neutrino properties by post-processing the neutrino signals extracted from no-oscillation CCSN simulations. The relative errors in neutrino number and energy luminosities of this phenomenological method are less than 2\% for 1D models, and less than 10\% for 2D models. This method provides a simple way to include the effects of FFC on neutrino signals without implementing a complex and expensive FFC scheme or redoing simulations.
\end{abstract} 

\ifMNRAS
\else 
\fi

\section{Introduction}
\label{sec:int}  

With the extremely high neutrino number density, the core of a core-collapse supernova (CCSN) serves as an ideal environment for collective neutrino oscillations \citep{duan2010,chakraborty2016,fischer2024,johns2025}. Fast flavor conversion (FFC), triggered by the fast flavor instability (FFI), is arguably the most discussed scenario in the collective oscillation context \citep{samuelsPhysRevD.48.1462,Sawyer_2005,volpe2015,Sawyer_2016,richers2019,johns2020,tamborra2021,padilla-gay_raffelt2022,richers_sen2022,morinaga2022,nagakura2023,nagakura2023b,volpe2024,xiong2024}. With timescales as short as a few nanoseconds, there is a possibility that FFC changes the neutrino radiation fields and signatures and the hydrodynamics of CCSN models in a significant way. There have been attempts to identify where in the supernova core angular crossings could occur \citep{abbar2019,glas_osc_2020,morinaga2020,abbar2021,johns_nagakura2021,nagakura_johns2021,richers2021,richers2021b,Cornelius2025}, but the feedback on supernova hydrodynamics is often missing in those studies. There have also been attempts to perform supernova and binary neutrino star (BNS) merger simulations with parameterized models of the effects of the FFC \citep{ehring2023a,ehring_help_or_hinder2023,mori2025_ffc,qiu2025,qiu2025b}. These generally make ad hoc assumptions about flavor mixing and where and when it occurs, and the results depend sensitively on the chosen free parameters whose values and appropriateness are quite uncertain. Therefore, the effects of FFC on the CCSN outcomes remain uncertain.

However, currently the full quantum kinetic calculation of the FFC is in general too expensive to be included in sophisticated radiation-hydrodynamic simulations of CCSNe. First, the instability criterion of the FFI is that the angular distribution of the difference between electron lepton number and heavy lepton number (ELN-XLN) crosses zero \citep{dasgupta2022,morinaga2022,Fiorillo2025}. To accurately capture the development of FFC, the neutrino angular distribution evolution needs to be closely followed, which requires an expensive multi-angle neutrino transport scheme. Compared to current multi-dimensional CCSN simulations, this may lead to a computational slow-down by factors of 10 to 20. Second, the timescale of the FFC can be as short as a few nanoseconds, which is about 500$-$1000 times shorter than the typical hydrodynamic timestep in CCSN simulations. In addition, as a quantum effect, the accurate treatment of FFC requires the evolution of the entire density matrix, instead of only its diagonal terms, which leads to another factor of two slow-down. In short, a comprehensive treatment of the FFC would make CCSN simulations four to five orders of magnitude slower, which will not be affordable any time soon.

Therefore, in order to make progress, several approximate methods have been explored.  By assuming that the angular distributions have a certain functional form, closure relations have been used in classic or quantum angular-moment methods to avoid the expensive multi-angle transport scheme \citep{Zhang_2013,johns2020,nagakura_johns2021,myers_richers_moments2022,froustey2024,kneller_QKE2024}. Instead of following the nanosecond-timescale evolution of the FFC, several methods have been proposed to predict the asymptotic/steady states of fast neutrino-flavor conversion \citep{richers2022,xiong_mengru2023,george2024,richers_box3D.2024,johns2025_thermodynamics,Goimil-Garcia2025}, which enables the use of a larger timestep in the simulations. Some of the asymptotic/steady state prediction methods rely on only the diagonal terms of the density matrix and can be applied to classic neutrino transport schemes. Therefore, each of the three major slow-down factors of the FFC calculation can be removed through reasonable approximations, and a combination of these methods allows the FFC to be included in CCSN simulations without undue extra computational cost.

\citet{wang2025} adopted such a combined method in 1D (spherical) and 2D (axis-symmetric) CCSN simulations and finds that the FFC effects significantly change the neutrino properties, while leaving CCSN hydrodynamics almost unchanged. This is because the FFC occurs mostly at relatively large radii where neutrino-matter interactions become very weak. Although the results look intriguing, this study has several limitations. First, \citet{wang2025} used a ``3-species'' neutrino transport scheme in which $\nu_\mu$, $\bar{\nu}_\mu$, $\nu_\tau$, and $\bar{\nu}_\tau$ were assumed to have exactly the same properties. Without FFC, this is a good approximation if muons and tauons are ignored. However, when the FFC is operating, differences between $\nu_e$ and $\bar{\nu}_e$ can be converted into differences between heavy lepton neutrinos and anti-neutrinos and enforcing $\nu_\mu=\bar{\nu}_\mu=\nu_\tau=\bar{\nu}_\tau$ breaks total ELN-XLN conservation. Although it is argued in \citet{wang2025} that the errors introduced by the ``3-species'' scheme are tolerable, such arguments need to be tested by a ``4-species'' scheme in which heavy lepton neutrinos and anti-neutrinos are treated separately. Second, the effects of the angular reconstruction methods, i.e., the closure choices, have yet to be studied. \citet{Cornelius2025} uses several different post-processing methods to identify the FFI unstable regions in 1D CCSN models and finds that different closures (and other types of ELN-XLN crossing identification methods) can lead to significantly different results. However, it's unclear if such sensitivity is due to the FFC itself or due to the different treatments of the neutrino angular distributions between the transport calculation and the post-processed FFI identification. In addition, it's unclear if feedback from the supernova simulation suppresses or enhances the sensitivity. Therefore, a self-consistent sensitivity study of angular reconstruction methods using an on-the-fly FFC treatment is preferred.

In this work, we present models in both 1D (spherical) and 2D (axisymmetric) done with different fast-flavor conversion schemes. We compare the FFC results calculated by the ``4-species'' and ``3-species'' schemes. Under the ``4-species'' scheme, we compare the FFC results calculated using three different closures. We find that the findings in \citet{wang2025} can be reproduced by all the FFC methods mentioned here, showing that the presented FFC effects are robust. In addition, we study in detail the FFC effects on the radiated supernova neutrinos. We find that the neutrino signals extracted from CCSN models without any FFC calculation can be post-processed by a ``quasi-equipartition'' method to mimic the FFC effects. This phenomenological method provides a simple way to include the effects of FFC on neutrino signals without implementing a complex and expensive FFC scheme and re-doing the simulations.

This paper is arranged as follows: in Section \ref{method}, we describe the details about the FFC schemes, angular reconstruction methods, and the CCSN models. We then present our results in Section \ref{results}. We compare different FFC supernova models in 1D and 2D, and analyze the FFC modulated neutrino signals. In Section \ref{conclusion}, we summarize our findings and discuss the caveats and limitations of this work.

\section{Method}
\label{method}
The CCSN simulations in this work are carried out using the sophisticated code (F{\sc{ornax}}), which has been described in detail in \citet{skinner2019}, \citet{burrows_40}, and in the appendix to \citet{vartanyan2019}. The neutrino-matter microphysics used in its classical physics sector can be found in \citet{2006NuPhA.777..356B} and \citet{2020PhRvD.102b3017W}. 
One major update to the F{\sc ornax} code in this study is that we adopt a ``4-species'' scheme instead of the commonly-used ``3-species'' scheme. In the ``3-species'' scheme, one ``$\mu$-type" neutrino type is used to represent all heavy neutrino and anti-neutrino species ($\nu_{\mu}$,$\bar{\nu}_{\mu}$,$\nu_{\tau}$, and $\bar{\nu}_{\tau}$), assuming they behave in exactly the same way. Although the neutrino-matter interaction cross-sections for such heavy neutrino species are very similar (ignoring the possible presence at high densities of muons), $\nu_e$ and $\bar{\nu}_e$ differences could result in differences between heavy-lepton flavor neutrinos and anti-neutrinos when neutrino flavor conversion is allowed. Since the ``3-species'' scheme always assumes $\nu_x=\bar{\nu}_x$, there is a caveat that such an assumption allows the indirect mixing between $\nu_e$ and $\bar{\nu}_e$ via $\nu_e\leftrightarrow\nu_x=\bar{\nu}_x\leftrightarrow\bar{\nu}_e$. Although it is argued in \citet{wang2025} that the errors introduced by the ``3-species'' scheme are tolerable, we here adopt a ``4-species'' scheme which distinguishes $\nu_x$ and $\bar{\nu}_x$ neutrios to fully account for this difference. In this work, heavy-lepton flavor neutrinos are assumed to have equal distributions ($\mathcal{F}_{{\nu}_\mu}=\mathcal{F}_{{\nu}_\tau}=\mathcal{F}_{{\nu}_x}$), and the same assumption is made for heavy-lepton flavor antineutrinos ($\mathcal{F}_{\bar{\nu}_\mu}=\mathcal{F}_{\bar{\nu}_\tau}=\mathcal{F}_{\bar{\nu}_x}$). 

In all these new simulations, we employ the SFHo nuclear equation of state \citep{2013ApJ...774...17S}, 1024 radial zones for the 1D simulations, and for the 2D simulations a grid of $1024\times128$ ($r\times\theta$). The outer boundary is set at 30,000 kilometers (km) and the inner radial zone is 0.5 km wide. The 9 M$_\odot$ progenitor model is taken from \citet{swbj16}, while the 18 M$_\odot$ is taken from \citet{sukhbold2018}. We use twelve neutrino energy groups from 1 to 300 MeV for each of four species ($\nu_e$, $\bar{\nu}_e$, $\nu_{x}$ and $\bar{\nu}_x$). 

To determine the effects of fast-flavor instability and conversion, F{\sc ornax} combines the Box3D oscillation formalism \citep{zaizen2023,richers_box3D.2024} with the BGK formalism \citep{Nagakura2024}. Although this scheme has already been described in \citet{wang2025}, we repeat the description here for reader's convenience.
Given the neutrino distribution $\mathcal{F}_{\nu_\alpha}(E,\hat{n})$ (where $E$ is the neutrino energy and $\hat{n}$ is the momentum direction vector), the $\alpha-$flavor lepton number angular distribution is (for simplicity we assume here $c=\hbar=1$):
\equ{
G_{\alpha}(\hat{n}) = \sqrt{2}G_F\int_{0}^{\infty}\frac{ {\rm d}E\,E^2}{(2\pi)^3}(\mathcal{F}_{\nu_\alpha}(E,\hat{n})-\mathcal{F}_{\bar{\nu}_\alpha}(E,\hat{n}))\, ,
}
and the ELN-XLN distribution in a given direction is
\equ{
G_{\hat{n}} = G_{e}(\hat{n}) - G_{x}(\hat{n})\, .
}

\begin{figure*}
    \centering
    \includegraphics[width=0.48\textwidth]{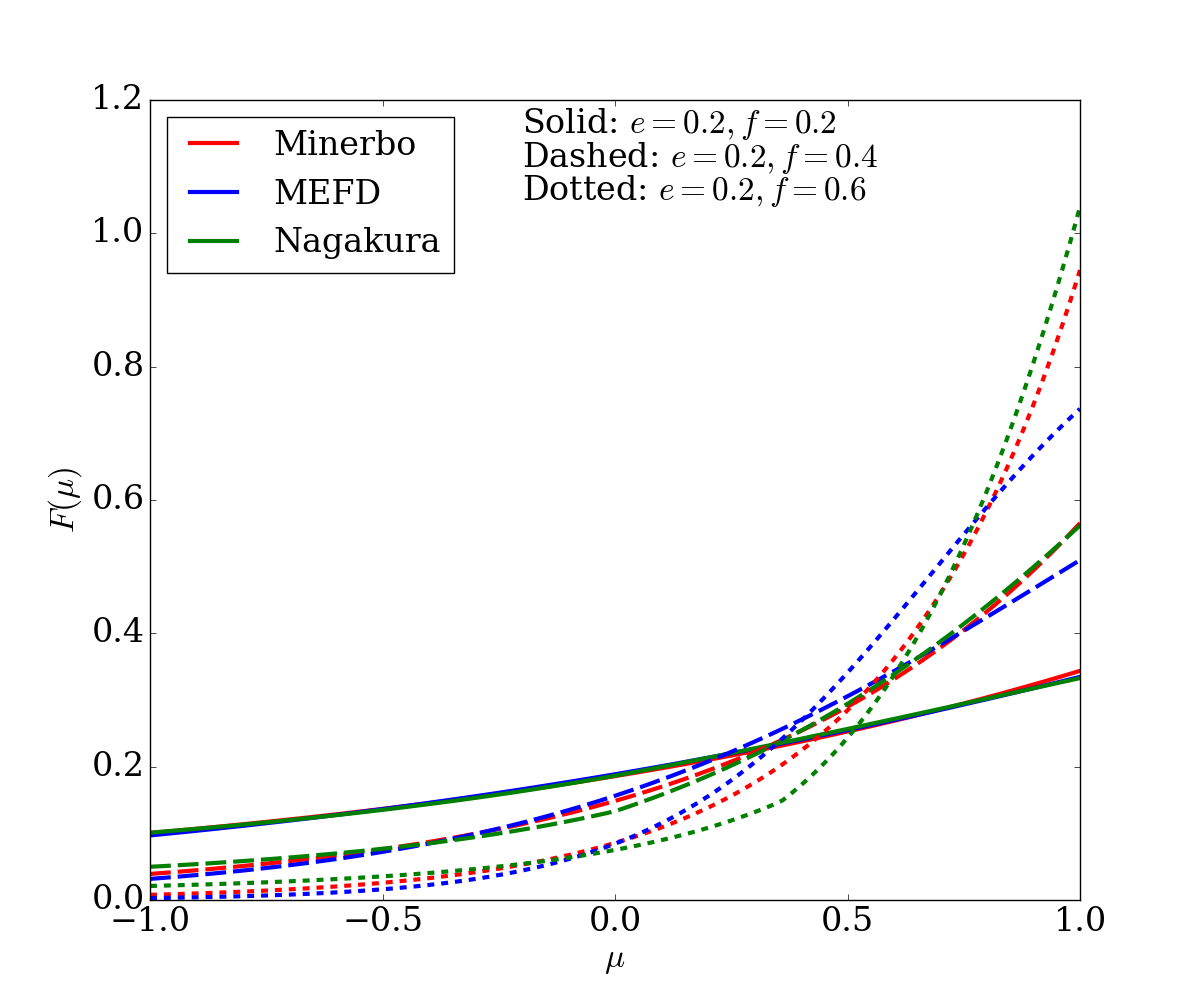}
    \caption{Reconstructed neutrino angular distributions of the Minerbo, MEFD, and Nagakura closures. With a flux factor ($f$) of 0.2, the differences between closure choices are almost negligible. At higher flux factors, the differences become more significant. One issue of the Minerbo and Nagakura closures is that they don't explicitly introduce the Pauli's exclusion principle, and as a result, their reconstructed angular distributions are more forwardly peaked than the MEFD results and the distribution function can even exceed one when the flux factor $f$ is close to $1-e$, where $e$ is the state occupation fraction. However, this issue leads only to very minor effects on the hydrodynamics of our models because neutrinos with high flux factors have already decoupled from matter.}
    \label{fig:closure}
\end{figure*}

The sign of $G$ divides the angular space into two regions: $\Gamma_+=\{\hat{n}\vert G(\hat{n})>0\}$ and $\Gamma_-=\{\hat{n}\vert G(\hat{n})<0\}$. With integrals defined on these two regions
\equ{
I_+=&\int_{\Gamma_+} {\rm d}\hat{n}G_{\hat{n}},\,\,\,\,I_-=-\int_{\Gamma_-} {\rm d}\hat{n}G_{\hat{n}}\, ,
}
the survival probability $P(\hat{n})$ can be expressed by
\equ{
P(\hat{n})=
\begin{cases}
      \frac{1}{3} &\,\, I_-<I_+, \hat{n}\in \Gamma_-, \\
      1-\frac{2I_-}{3I_+} &\,\, I_-<I_+, \hat{n}\in \Gamma_+, \\
      \frac{1}{3} &\,\, I_->I_+, \hat{n}\in \Gamma_+, \\
      1-\frac{2I_+}{3I_-} &\,\, I_->I_+, \hat{n}\in \Gamma_-, \\
   \end{cases}
}
and the asymptotic angular distributions are given by
\equ{
\mathcal{F}^{\rm FFC}_{\nu_e}(E,\hat{n})=&P(\hat{n})\mathcal{F}_{\nu_e}(E,\hat{n})+[1-P(\hat{n})] \mathcal{F}_{\nu_x}(E,\hat{n}),\\
\mathcal{F}^{\rm FFC}_{\bar{\nu}_e}(E,\hat{n})=&P(\hat{n})\mathcal{F}_{\bar{\nu}_e}(E,\hat{n})+[1-P(\hat{n})] \mathcal{F}_{\bar{\nu}_x}(E,\hat{n}),\\
\mathcal{F}^{\rm FFC}_{\nu_x}(E,\hat{n})=&\frac{1}{2}[1-P(\hat{n})]\mathcal{F}_{\nu_e}(E,\hat{n})+\frac{1}{2}[1+P(\hat{n})] \mathcal{F}_{\nu_x}(E,\hat{n}),\\
\mathcal{F}^{\rm FFC}_{\bar{\nu}_x}(E,\hat{n})=&\frac{1}{2}[1-P(\hat{n})]\mathcal{F}_{\bar{\nu}_e}(E,\hat{n})+\frac{1}{2}[1+P(\hat{n})] \mathcal{F}_{\bar{\nu}_x}(E,\hat{n})\,.\\
\label{eq:formula}
}
The following formula is used to estimate the local growth rate of fast flavor instability \citep{morinaga2020,Nagakura2024}:
\equ{
\sigma = \sqrt{I_+I_-}\, ,
\label{eq:growth}
}
and the change in the neutrino angular distribution per simulation timestep $\Delta t$ is handled by the BGK scheme: \equ{
\mathcal{F}'_{\nu_\alpha} - \mathcal{F}_{\nu_\alpha} = -(1-e^{-\sigma \Delta t})(\mathcal{F}_{\nu_\alpha}-\mathcal{F}^{\rm FFC}_{\nu_\alpha})\, .
\label{control}}
The angular moments used by the M1 transport scheme are calculated from such distributions via numerical integrations in angular space. With the relaxation time defined by the inverse growth rate, we apply the scheme to the entire simulation domain and let eq. \ref{control} handle the ``freeze-out" of the flavor conversion at large radii where growth rates are low. Since Eq.\ref{eq:growth} is not an exact formula for FFI growth rates, some uncertainties have been introduced. However, our conclusions remain unaltered, even if we vary the growth rate $\sigma$ by a factor of 5 or 1/5. This is because in most regions where the FFC is important, the conversion timescale is short enough that the ELN-XLN crossing is erased instantly every timestep. Therefore, the actual bottleneck of flavor conversion is not the timescale, but the degree of ELN-XLN crossing. The uncertainty in the FFI growth rates becomes important only at large radii where growth timescales are comparable to the hydrodynamic timesteps in the simulations, but FFC at such radii is in this instance weak.

It is worth mentioning that the numerical uncertainties of such integrations become larger when the neutrino angular distribution becomes very forward-peaked at large radii. This is due to the fact that although we use Lebedev quadrature on the sphere with 110 points to perform the necessary angular integrations, integrating exactly all spherical harmonics up to 17th order, the integration scheme cannot resolve the Dirac-delta-function-like behavior of the very forward-peaked neutrino angular distributions. However, we argue that such numerical uncertainties will not affect our results and the conclusions of this work because they occur only at large radii, where local growth rates and conversion fractions are low.

\begin{figure*}
    \centering
    \includegraphics[width=0.48\textwidth]{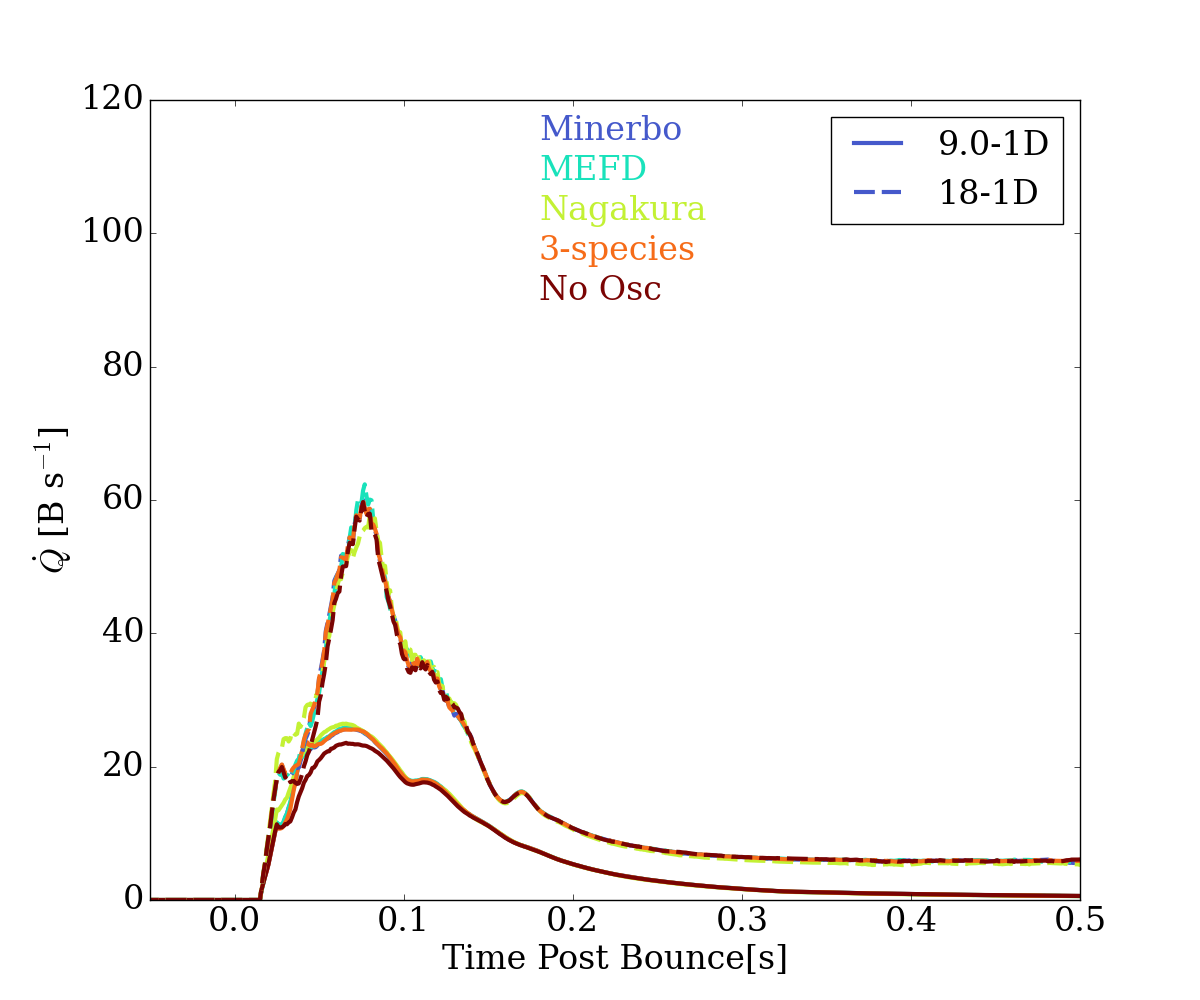}
    \includegraphics[width=0.48\textwidth]{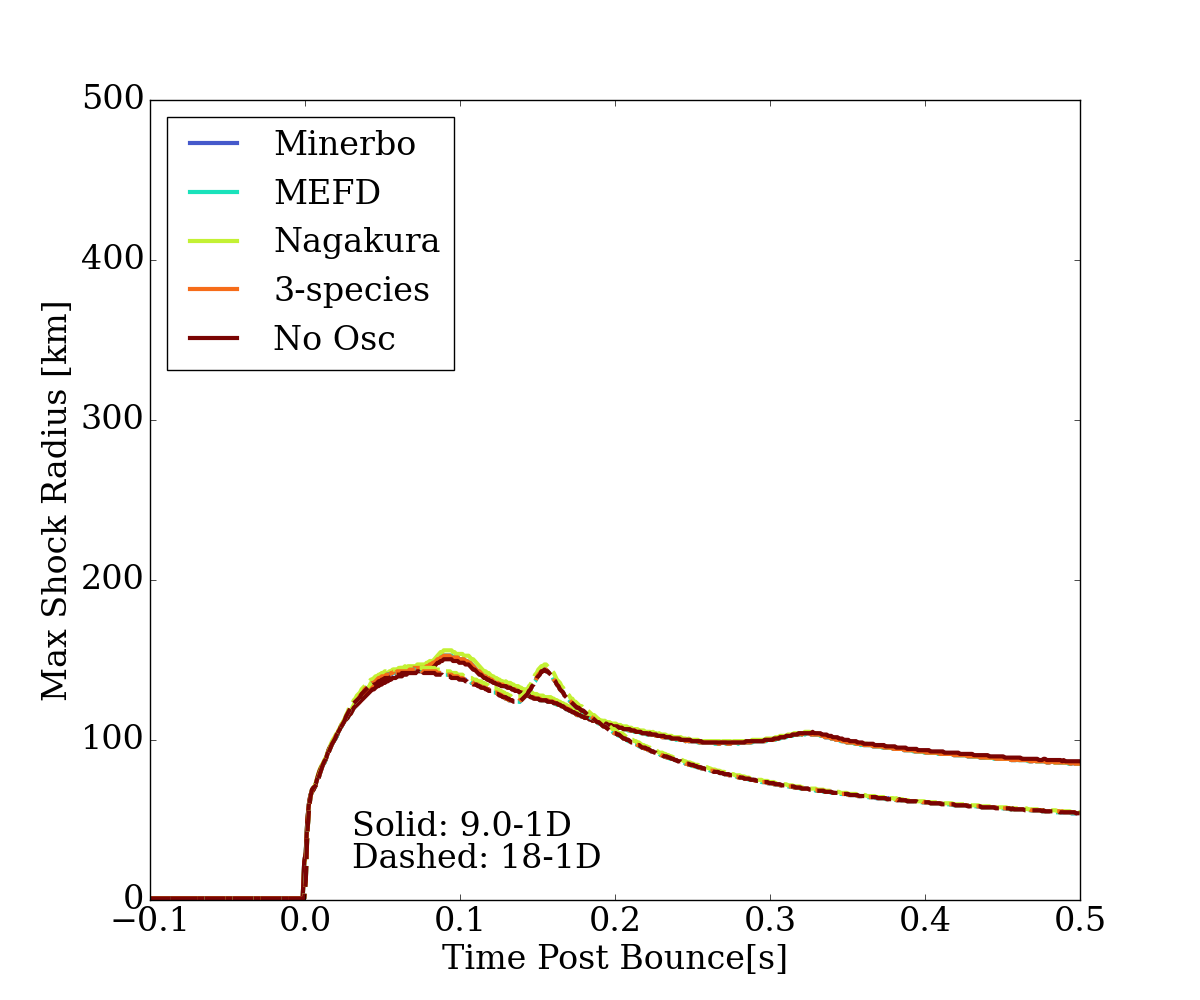}
    \caption{Left: Net neutrino heating rate (in unit of $10^{51}$ erg $s^{-1}$) in the gain region behind the shock for the 1D models. Right: the maximum shock radius as a function of time. The 9 M$_\odot$ model is shown by solid lines, while the 18 M$_\odot$ model is marked by dashed lines. Different colors indicate the different treatments of neutrino fast flavor conversion. Models using the 4-species scheme and various angular reconstruction methods (``Minerbo'', ``MEFD'', and ``Nagakura'') all behave very similar to the 3-species scheme model (``3-species''). Compared to the no oscillation models (``No osc''), models with FFC show slightly higher heating rates at early time in the 9 M$_\odot$ model, while in general the effects of FFC on heating rates and shock evolution are weak. This confirms the findings in \citet{wang2025}. }
    \label{fig:heating}
\end{figure*}

\subsection{Angular Reconstruction Methods}
Since F{\sc ornax} uses an M1 neutrino transport scheme \citep{skinner2019} which evolves only the zeroth and first moments of neutrino angular distributions, an angle-dependent closure relation is needed to reconstruct the full angular distribution $\mathcal{F}_{\nu_\alpha}(E,\hat{n})$ and to derive the higher order angular moments from the normalized zeroth and first moments -- the occupation fraction $e$, and the flux factor $f$. The relation between $e$ and $f$ and the angular distribution function $\mathcal{F}(\hat{n})$ is $e=\int d\hat{n}\mathcal{F}(\hat{n})$ and $f=\frac{1}{e}\int d\hat{n}\mathcal{F}(\hat{n})\hat{n}$. Furthermore, we assume that the neutrino angular distributions are axisymmetric, i.e., $\mathcal{F}_{\nu_\alpha}(E,\hat{n})=\mathcal{F}_{\nu_\alpha}(E,\mu)$, where $\mu=\hat{n}\cdot\hat{f}$ and $\hat{f}$ is the neutrino flux direction.

Seven closures were listed in \citet{wang_burrows2023} to study the impact of different closure choices on CCSN simulations without oscillation effects. However, since many closure relations don't provide the full angular distribution (only the the second and third angular moments), only three out of these seven closures can be used in our Box3D-BGK scheme:
\begin{itemize}
    \item The Minerbo closure: $\mathcal{F}(\mu) = \exp(-\eta+\alpha\mu)$
    \item The maximum entropy Fermi-Dirac (MEFD) closure: $\mathcal{F}(\mu) = (1+\exp(\eta-\alpha\mu))^{-1}$
    \item The Nagakura closure: 
$
\ln \mathcal{F}(\mu)= 
\left\{
\begin{array}{c}
a\mu^2+b\mu+c \,\,\,\,(\mu>\mu_0)\\
d\mu^2+g\mu+h \,\,\,\,(\mu<\mu_0)\\
\end{array}
\right.\nonumber
$
\end{itemize}
where $\alpha$, $\eta$, $a$, $b$, $c$, $d$, $g$, $h$, and $\mu_0$ are parameters. The two parameters $\alpha$ and $\eta$ in the Minerbo and MEFD closures are numerically solved using $e$ and $f$, while the seven parameters in the Nagakura closure are interpolated from the table provided by \citet{nagakura2021_closure}. A comparison of these closures under several sets of ($e,f$) is shown in Figure \ref{fig:closure}. The differences between closures are more significant with higher flux factors, especially when $f$ is close to its upper limit $1-e$. One issue of the Minerbo and Nagakura closures is that they don't explicitly enforce the Pauli exclusion principle, and as a result, their reconstructed angular distributions can be more forward peaked than the actual distribution and can even exceed one. However, we think this issue leads to only very minor errors since high flux factors can be achieved only at large radii where the FFC is weak. In addition, since the angular distribution serves only as the auxiliary state to calculate the amount of energy and flux converted by the FFC, the unphysical behavior will not get amplified in the other parts of the code. 

\section{Results}  
\label{results}

We have calculated for this study models in both 1D (spherical) and 2D (axisymmetric) with different fast-flavor conversion schemes. We select two representative progenitors: the 9 M$_\odot$ model from \citet{swbj16} and the 18 M$_\odot$ model from \citet{sukhbold2018}. For each progenitor, we calculate three FFC models using the ``4-species'' scheme with the Minerbo, MEFD, and Nagakura closures, one model using the ``3-species'' scheme with the Minerbo closure, and one model without any flavor conversion. The motivations for such comparison are (a) to study the impact of different angular reconstruction methods, and (b) to study the difference between the ``4-species'' and ``3-species'' schemes. 

\begin{figure*}
    \centering
    \includegraphics[width=0.48\textwidth]{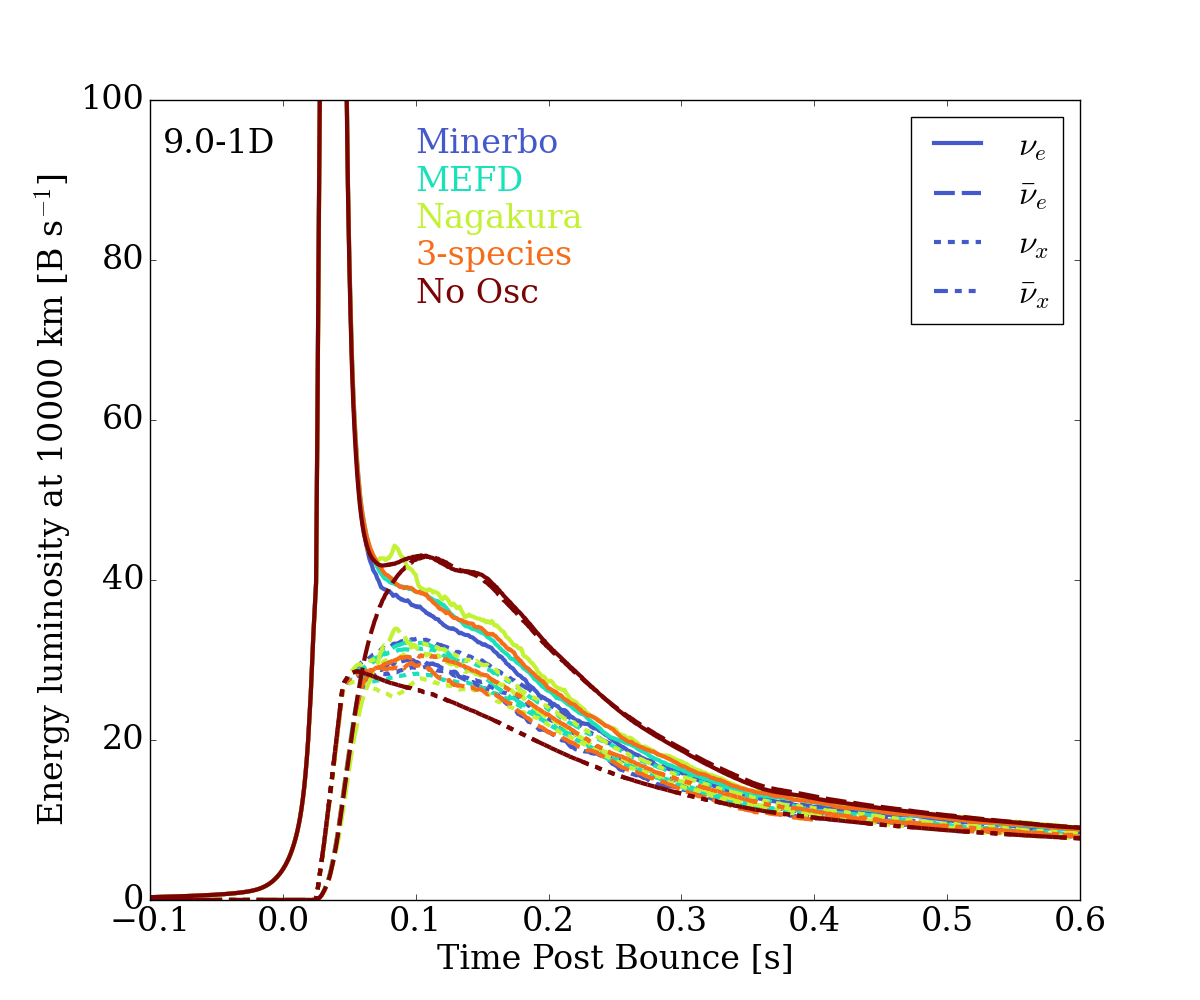}
    \includegraphics[width=0.48\textwidth]{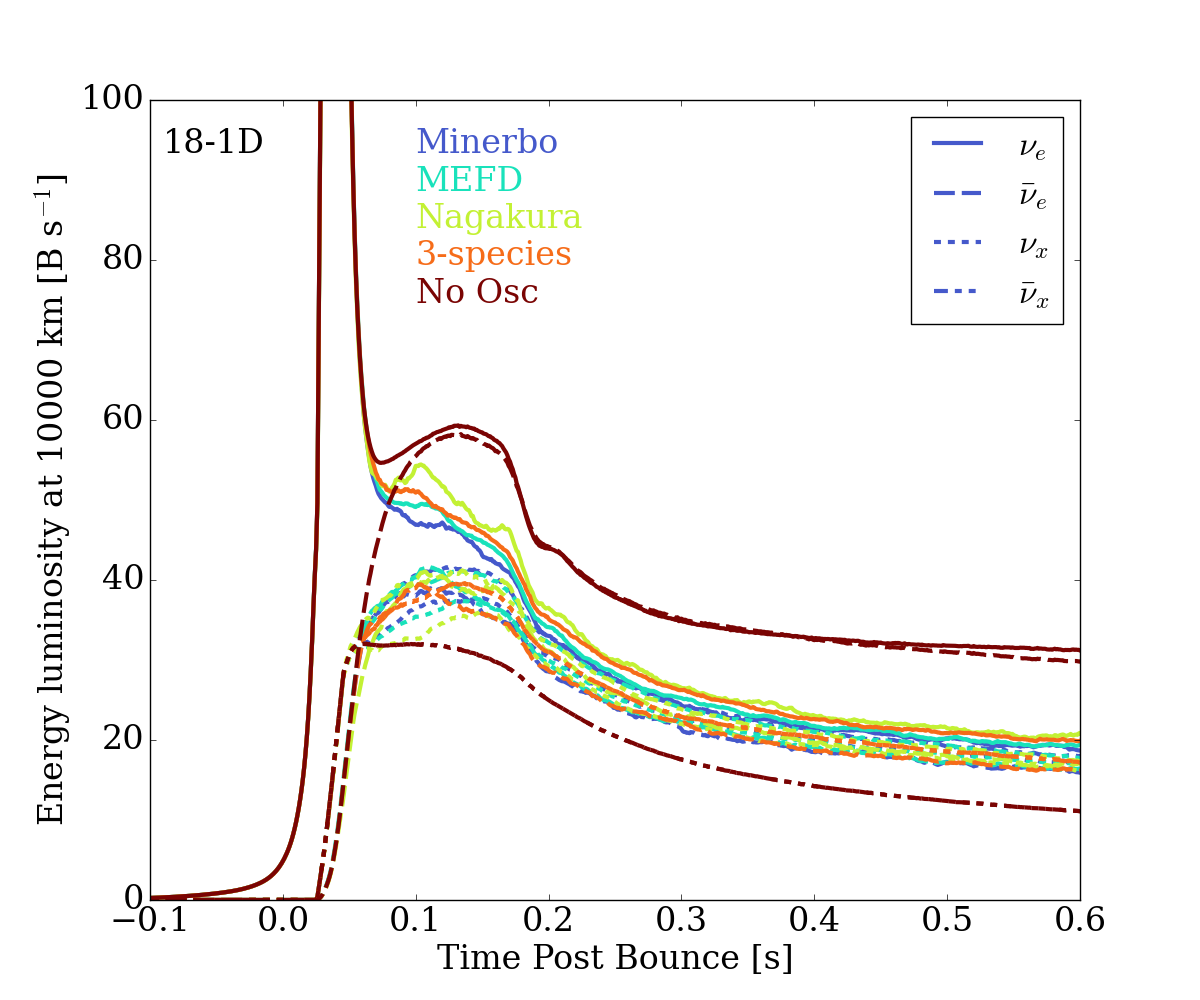}
    \caption{Neutrino luminosities (in unit of $10^{51}$ erg s$^{-1}$) as a function of time for the 1D models. Different line styles indicate different neutrino types. For the ``3-species'' and ``No osc'' models, for which the 3-species scheme is used, $\nu_x$ and $\bar{\nu}_x$ have exactly the same luminosities and are thus overlapping. After a slight delay, fast-flavor conversion boosts the $\nu_{x}$ and $\bar{\nu}_{x}$ luminosities by $\sim$20\%. Differences between various angular reconstruction methods and FFC schemes are relatively minor. Without the FFC, the $\nu_e$ and $\bar{\nu}_e$ have similar energy luminosities which are significantly higher than those of $\nu_x$ and $\bar{\nu}_x$ neutrinos. This trend is completely changed by flavor conversion: for models with FFC, the relative differences between energy luminosities of various neutrino are less than about 10\% after $\sim200$ ms post-bounce, and the luminosity order at that time is $L_{\nu_e}>L_{\bar{\nu}_x}>L_{\nu_x}>L_{\bar{\nu}_e}$. }
    \label{fig:luminosities}
\end{figure*}

\begin{figure*}
    \centering
    \includegraphics[width=0.48\textwidth]{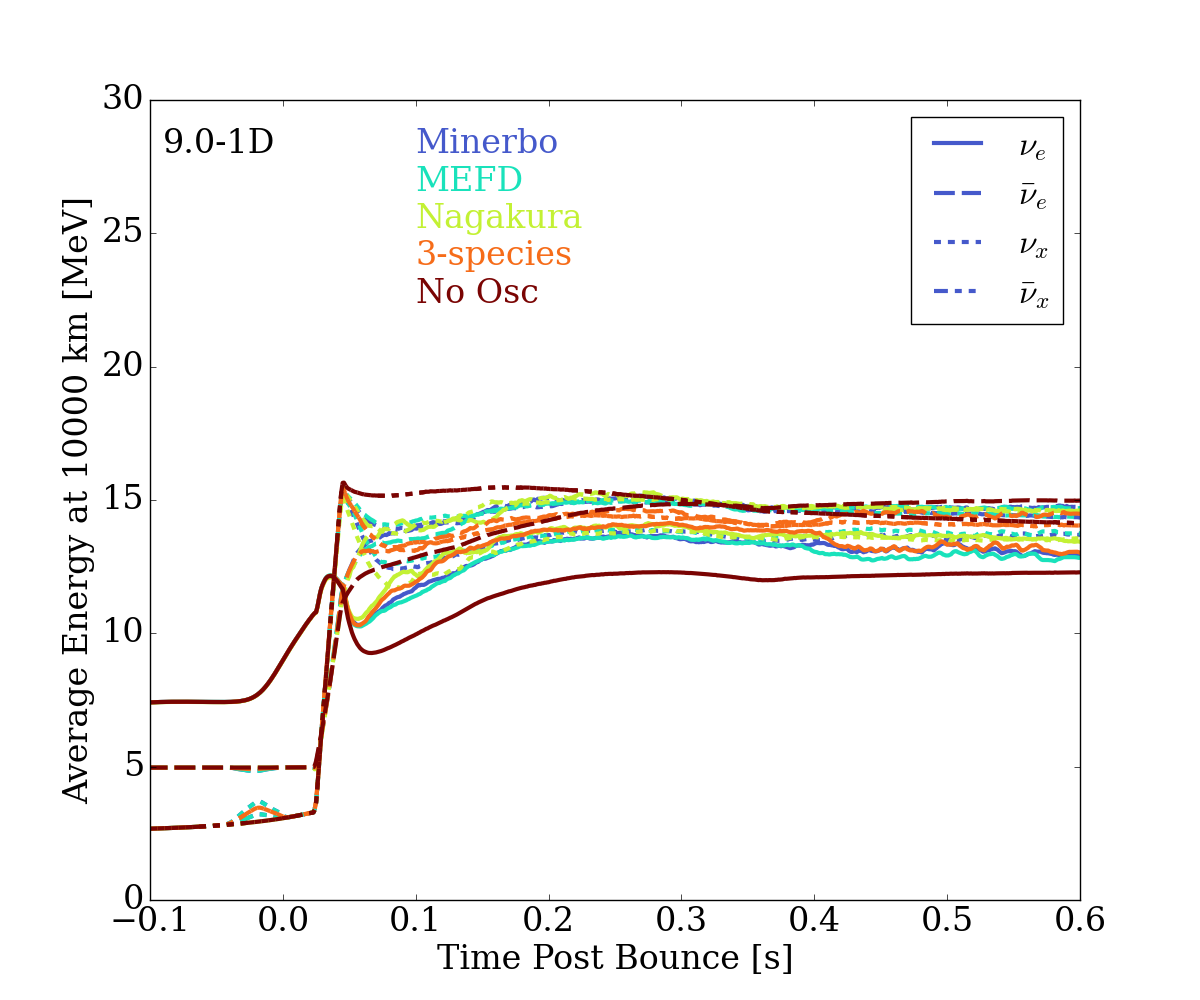}
    \includegraphics[width=0.48\textwidth]{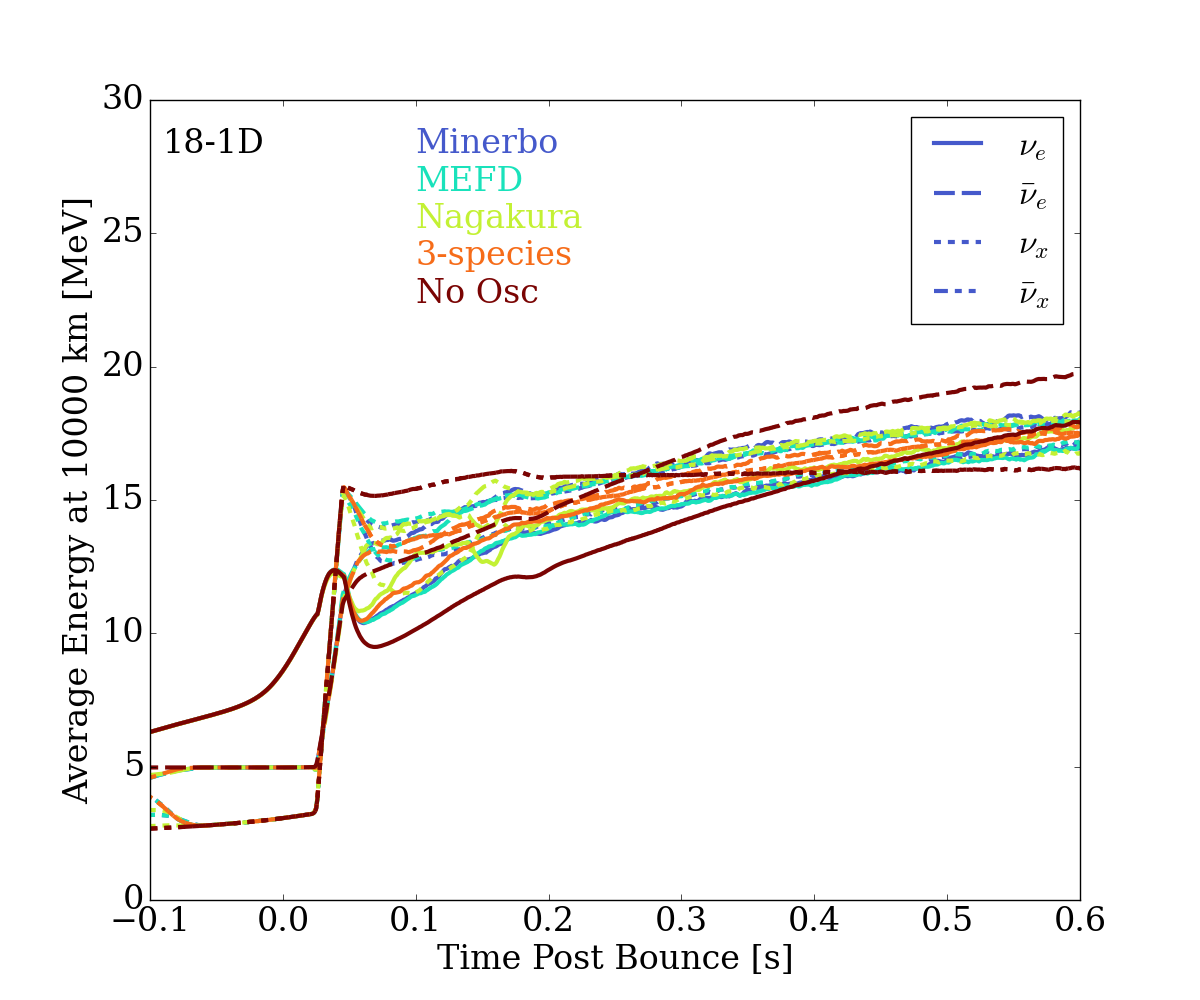}
    \caption{The evolution of the average energies of the different neutrino types. Different line styles indicate different neutrino types. When the FFC is operative, the $\nu_{x}$ and $\bar{\nu}_{x}$ neutrino spectra are softened, while both the $\nu_e$ and $\bar{\nu}_e$ neutrino spectra harden slightly. The differences between various angular reconstruction methods are minor. With FFC, the average energies of the four neutrino types show smaller, but non-vanishing, differences.  Neutrinos ($\nu_e$ and $\nu_x$) have similar average energies, while anti-neutrinos ($\bar{\nu}_e$ and $\bar{\nu}_x$) have similar average energies, but at higher values. } 
    \label{fig:spectra}
\end{figure*}

\begin{figure*}
    \centering
    \includegraphics[width=0.48\textwidth]{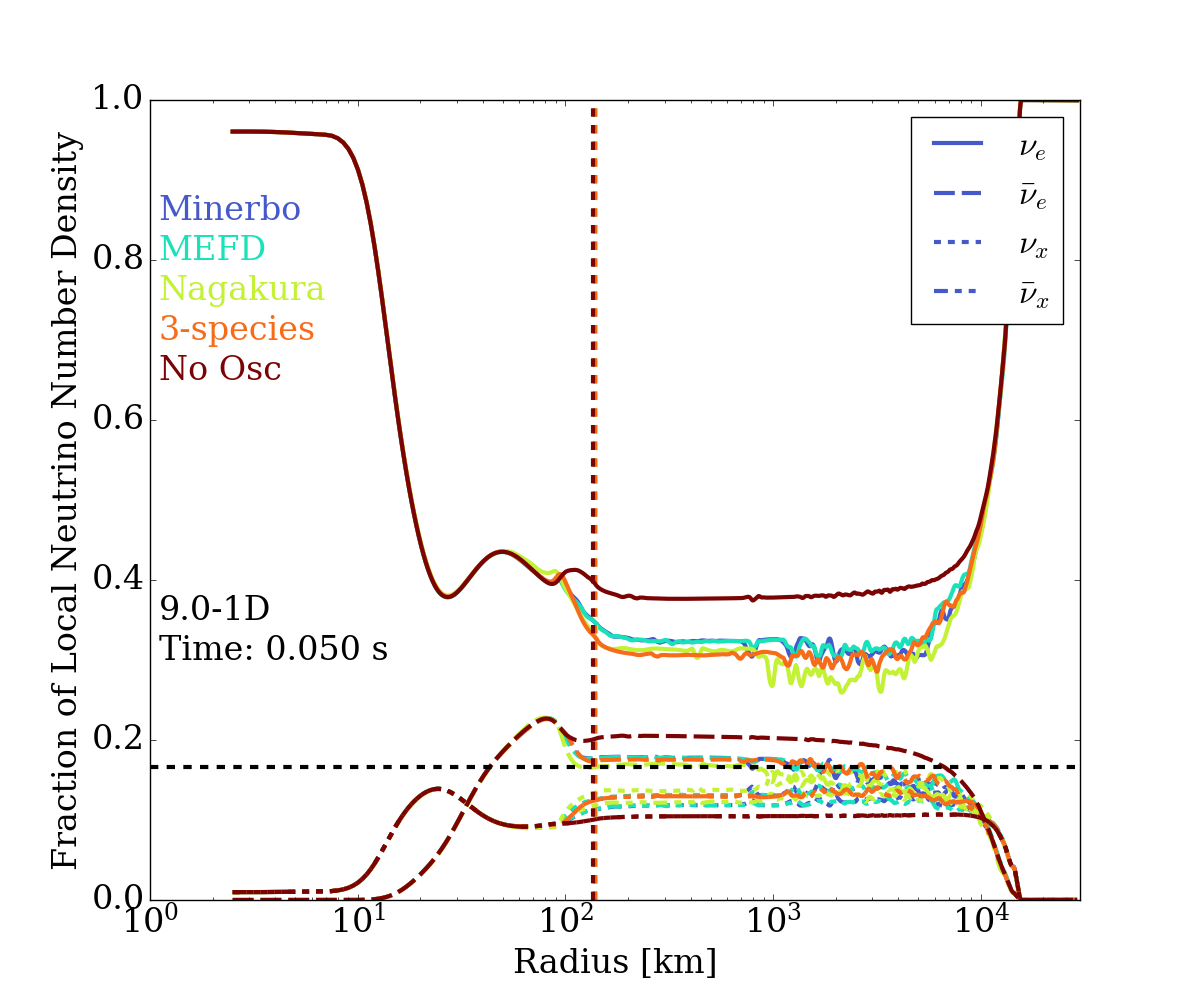}
    \includegraphics[width=0.48\textwidth]{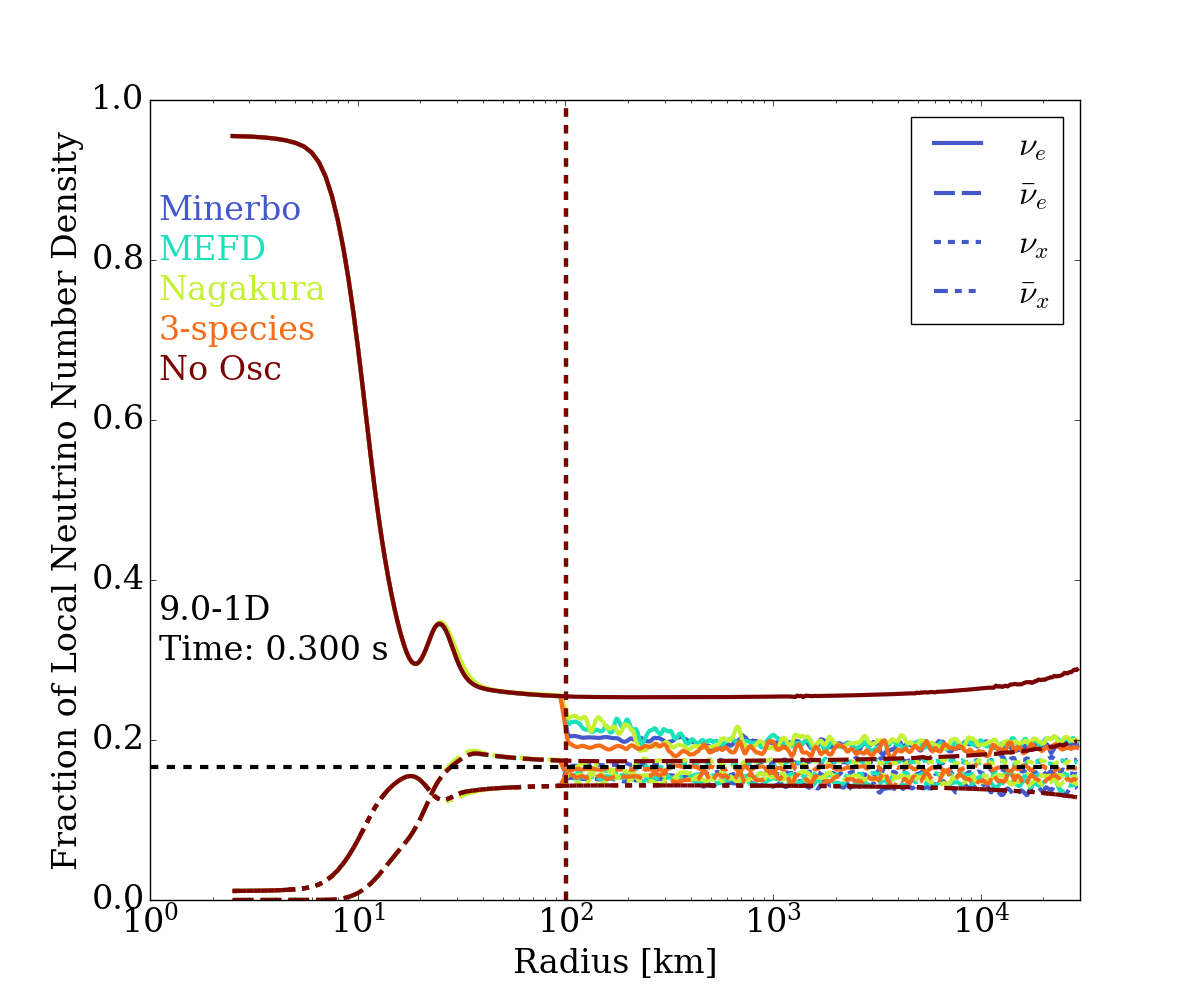}
    \includegraphics[width=0.48\textwidth]{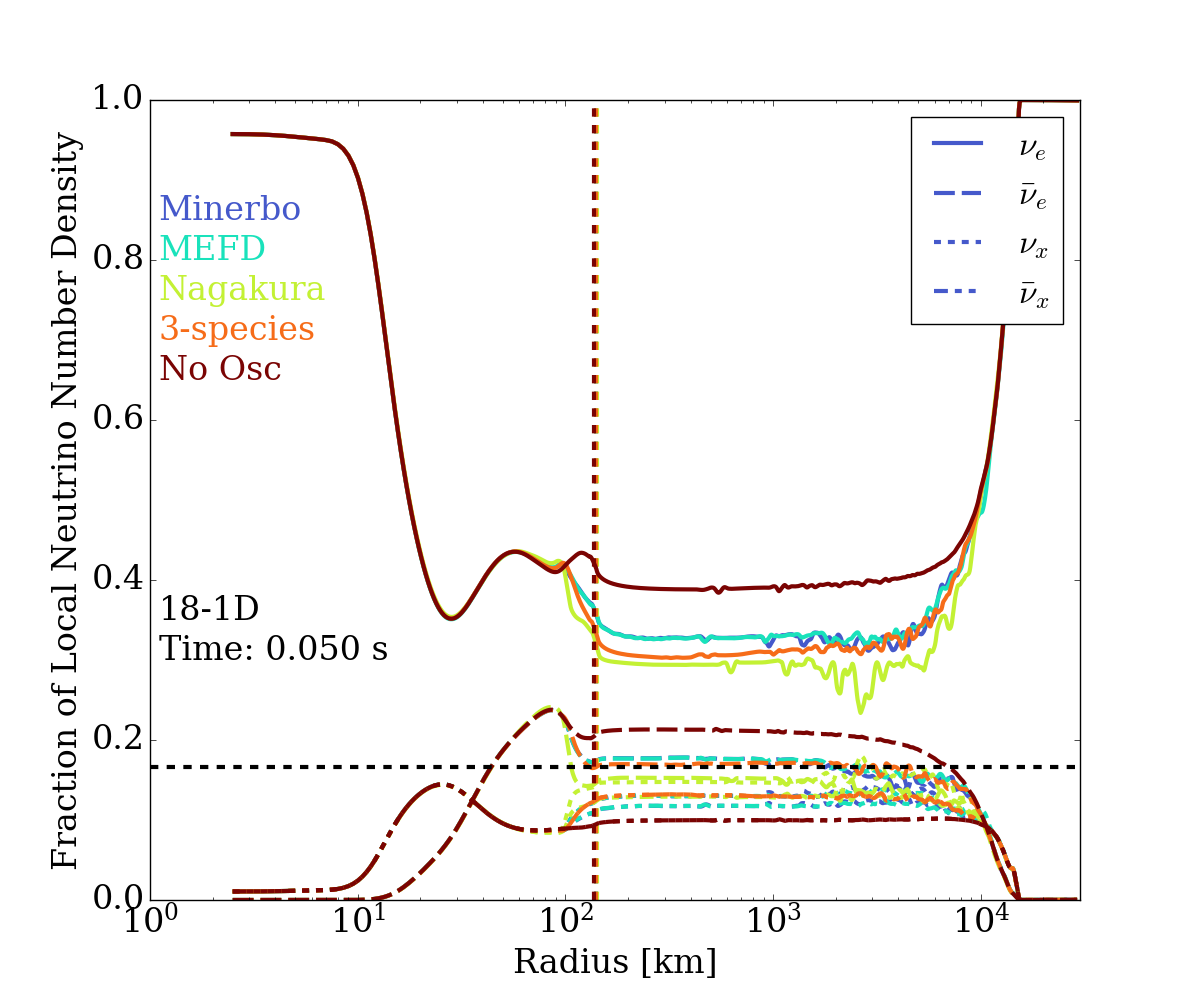}
    \includegraphics[width=0.48\textwidth]{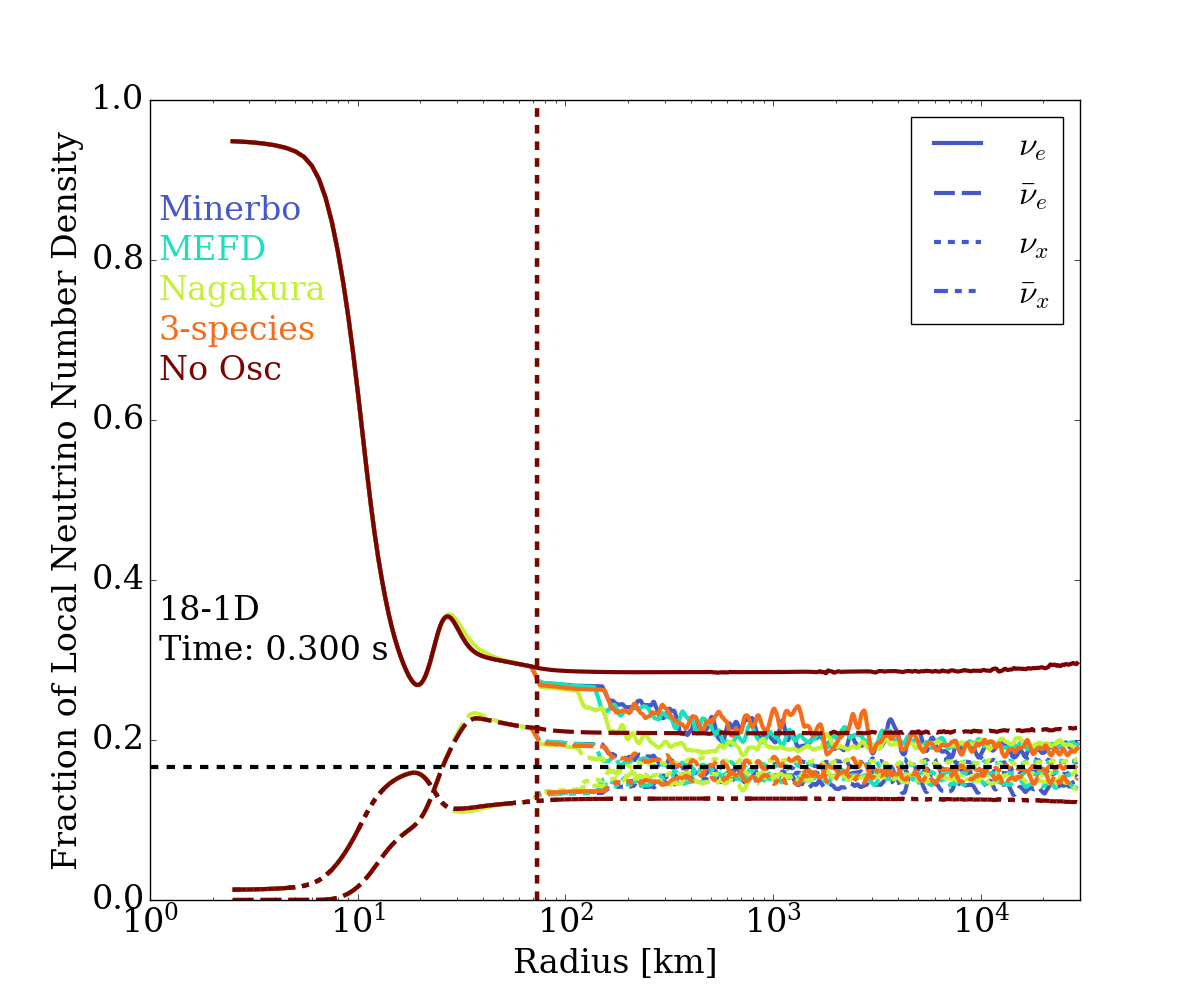}
    \caption{The neutrino number density fraction profiles for the 1D models at 50 and 300 ms post-bounce. The vertical dashed line marks the location of the shock wave, while the horizontal dashed line at $1/6\approx0.167$ marks the equipartition fraction. At early times (50 ms post-bounce), the FFC happens mostly in the FFI region interior to the shock. At relatively later times ($\sim$300 ms post-bounce), the post-shock FFI region disappears and all flavor conversion happens exterior to the shock. The final flavor state is gradually approached at thousands of kilometers.}
    \label{fig:fractions}
\end{figure*}

\subsection{One-Dimensional Models}

Figure \ref{fig:heating} shows the net neutrino heating rates and maximum shock radius evolution for the 1D models. Models using the 4-species scheme and various angular reconstruction methods (“Minerbo”, “MEFD”, and “Nagakura”) all behave very similarly to models using the 3-species scheme. After a short delay post-bounce, the net heating rates in the 9 M$_\odot$ model are enhanced by about 10\% for a few tens of milliseconds and then settle to that without the FFC effects. In the 18 M$_\odot$ model, no significant FFC effects are seen, except that the heating rate in the model using Nagakura closure is enhanced for about 20 ms. Such a short period of heating rate enhancement is insufficient to change the hydrodynamic behavior of the model, as shown by the shock radius evolution. The shock radii in all models, with or without FFC effects, deviate only by no more than 3 kilometers. The deviation caused by the FFC effects is even smaller, as the impact of closure choices on classic neutrino transport can also introduce kilometer-level changes to the shock position \citep{wang_burrows2023}.

The evolution of the neutrino luminosities measured at 10000 km for the 1D models is shown in Figure \ref{fig:luminosities}. The effects of FFC start at about 50 ms post-bounce and last throughout the simulations. Compared to the no oscillation models, about 20\% of the electron and anti-electron neutrino luminosities are converted by the FFC into the $x$- and anti-$x$-type neutrino luminosities. This conversion fraction decreases as a function time in the 9 M$_\odot$ models, while in the 18 M$_\odot$ models it can increase to about 50\% at later times. The different behaviors are because the 18 M$\odot$ models experience significantly stronger accretion, which powers the luminosities and strengthens flavor conversion.

Without the FFC, the $\nu_e$ and $\bar{\nu}_e$ have similar energy luminosities, which are significantly higher than that of the $\nu_x$ and $\bar{\nu}_x$ neutrinos individually. This trend is completely changed by flavor conversion: for models with FFC, the relative differences between energy luminosities of various neutrinos are less than about 10\% after $\sim200$ ms post-bounce, and the luminosity order at that time is $L_{\nu_e}>L_{\bar{\nu}_x}>L_{\nu_x}>L_{\bar{\nu}_e}$. 

\begin{figure*}
    \centering
    \includegraphics[width=0.48\textwidth]{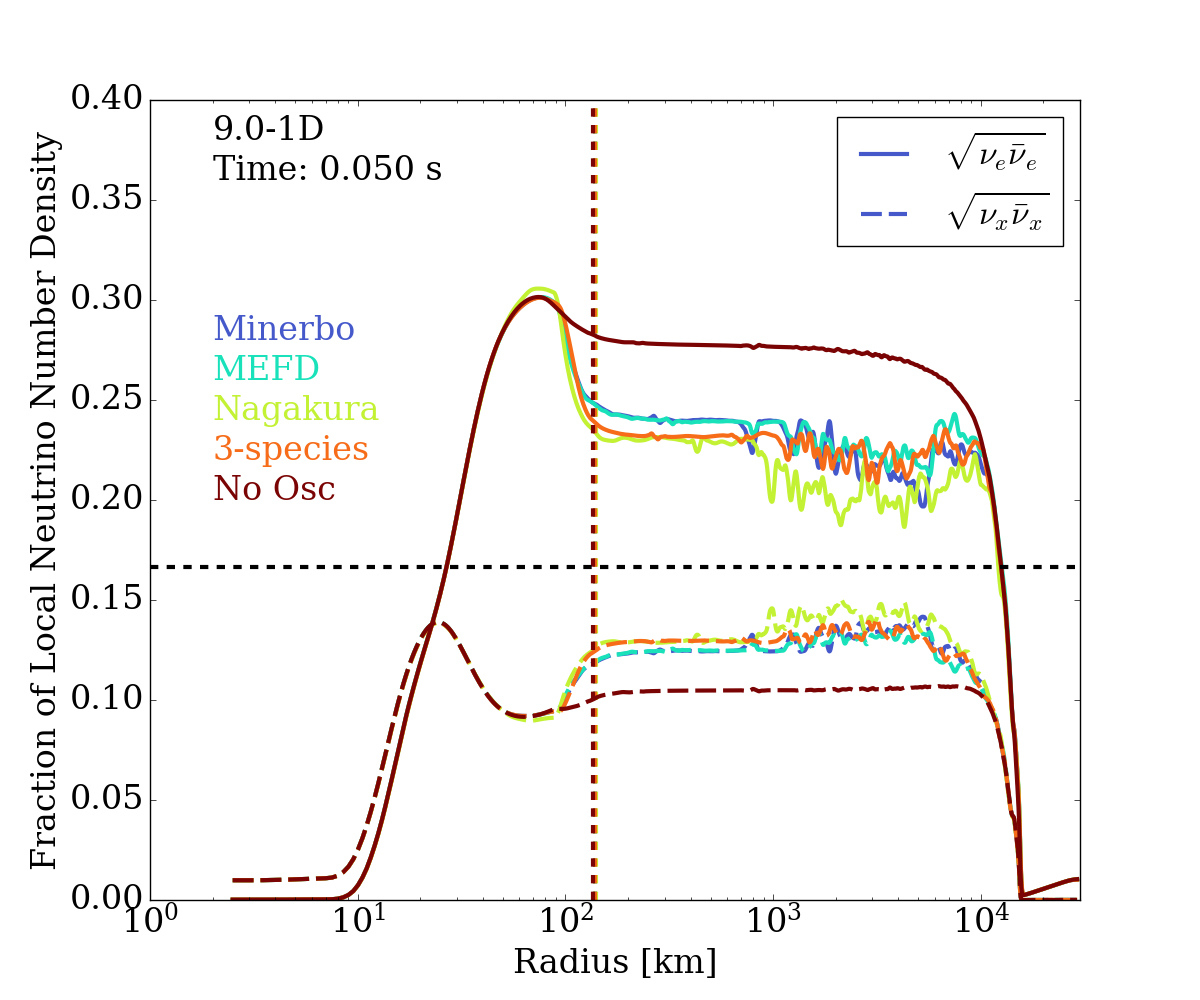}
    \includegraphics[width=0.48\textwidth]{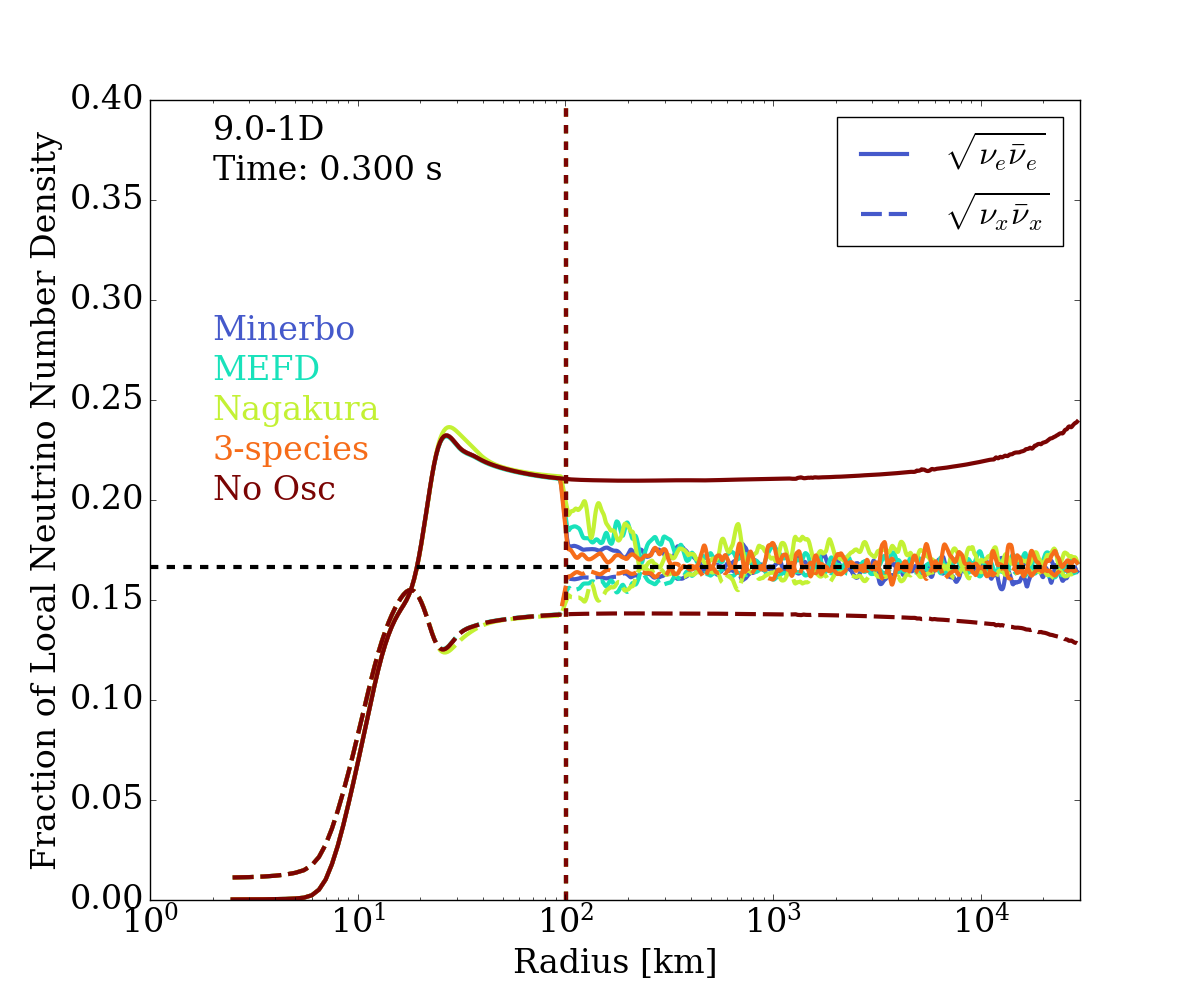}
    \includegraphics[width=0.48\textwidth]{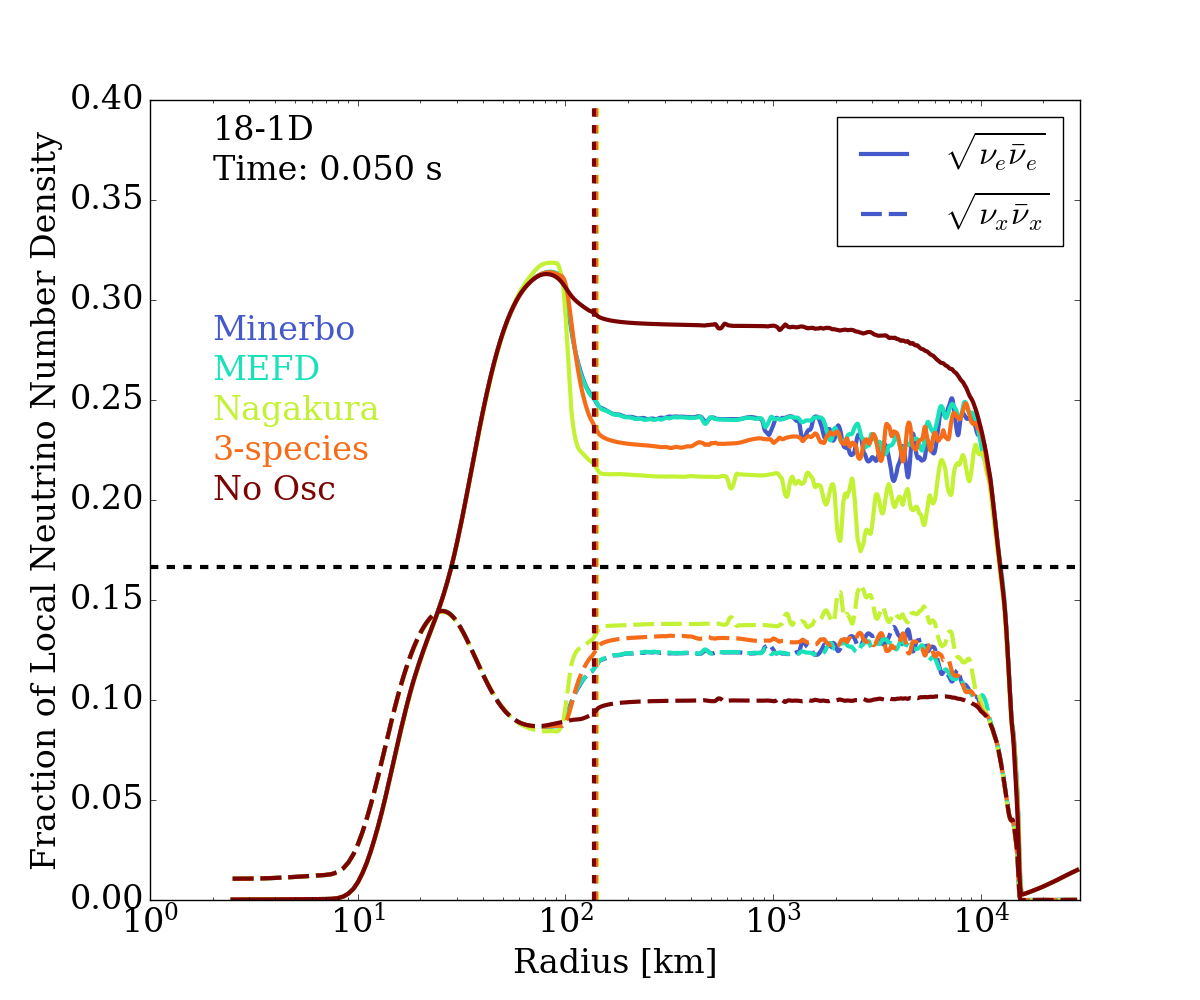}
    \includegraphics[width=0.48\textwidth]{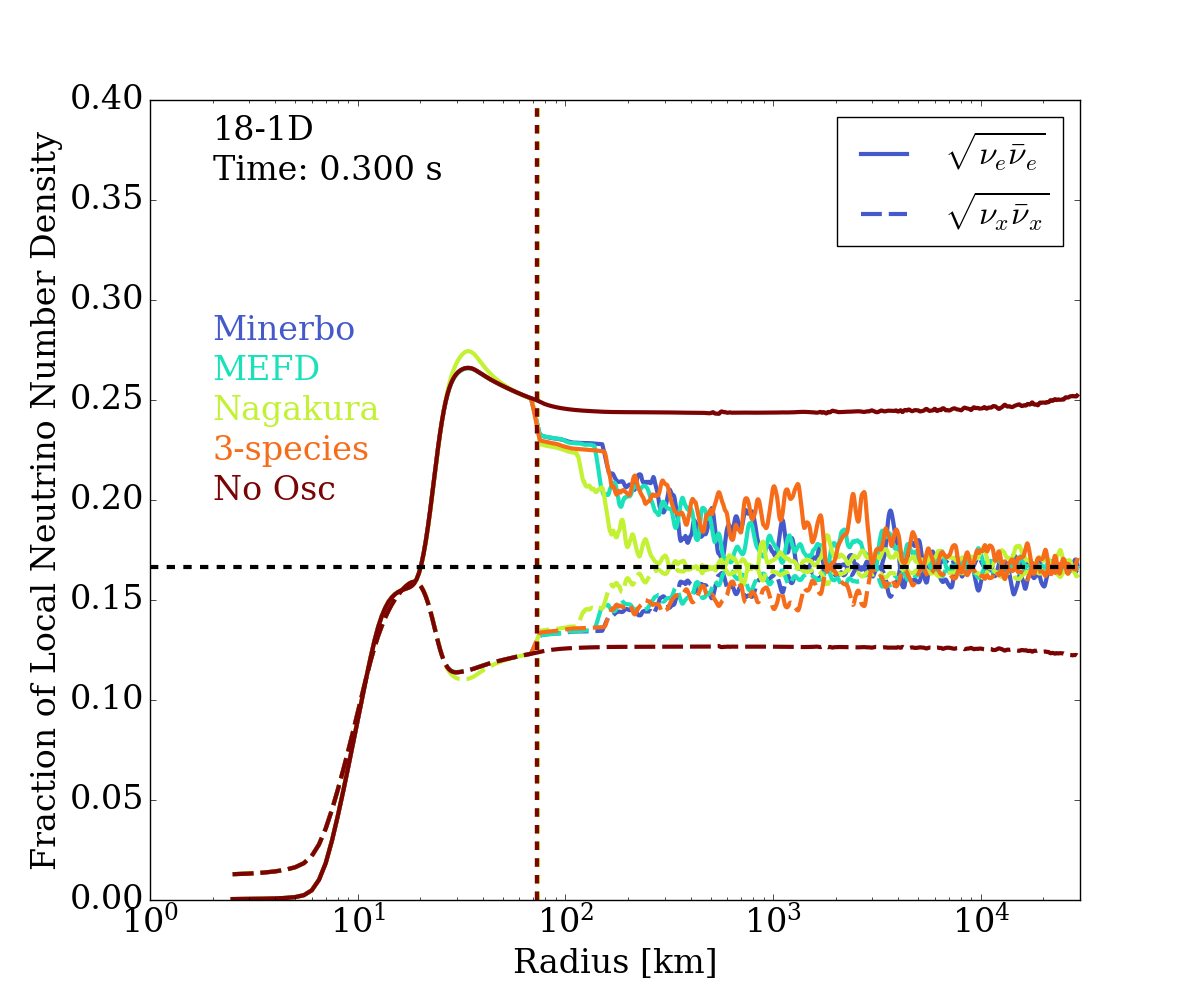}
    \caption{The neutrino number density fraction profiles of $\sqrt{\nu_e\bar{\nu}_e}$ and $\sqrt{\nu_x\bar{\nu}_x}$ for the 1D models at 50 and 300 ms post-bounce. The vertical dashed line marks the location of the shock wave, while the horizontal dashed line at $1/6\approx0.167$ marks the equipartition fraction. Although at early times ($\sim$50 ms) the neutrino number density fraction doesn't satisfy $\sqrt{N^{\rm FFC}_{\nu_e}N^{\rm FFC}_{\bar{\nu}_e}}=\sqrt{N^{\rm FFC}_{\nu_x}N^{\rm FFC}_{\bar{\nu}_x}}$, this quasi-equipartition condition serves as a good approximation at later times. }
    \label{fig:asymptotic}
\end{figure*}

Figure \ref{fig:spectra} shows the evolution of the average energies of different neutrino types for the 1D models. All FFC models show similar behavior regardless of the exact implementation of the FFC scheme. When the FFC is operative, the $\nu_{x}$ and $\bar{\nu}_{x}$ neutrino spectra are softened, while both the $\nu_e$ and $\bar{\nu}_e$ neutrino spectra harden slightly.  With FFC, the average energies of the four neutrino types show smaller, but non-vanishing, differences. Neutrinos ($\nu_e$ and $\nu_x$) have similar average energies, while anti-neutrinos ($\bar{\nu}_e$ and $\bar{\nu}_x$) have similar average energies, but at higher values. 

Figure \ref{fig:fractions} shows the neutrino number density fraction profiles for the 1D models at two selected snapshots. Such fraction profiles provide a better way to show where the FFC is happening than the FFI growth rate profiles. First, the FFI growth rates don't reflect the actual amount of flavor conversion. While the growth rates are high, flavor conversion might be very weak because the conversion stops as soon as the ELN-XLN crossing is erased. Second, the growth rates are dependent on the hydrodynamic timesteps of our simulations. Since the ELN-XLN crossing from previous timesteps will be erased after the calculation of the FFC, the degree of ELN-XLN crossing which is used to compute growth rates scales with the current timestep duration. This numerical behavior has little impact on our results, other than the growth rates, because such issues only occur where the FFI growth timescale is shorter than the hydrodynamic timestep, and flavor conversion will happen instantly even without overestimating the growth rate.

In Figure \ref{fig:fractions}, we see that the major FFC region changes with time. Shortly after core bounce, the FFC happens mostly in the outer neutrino gain region interior to the shock. Although the FFI growth timescales in such regions are significantly shorter than the hydrodynamic timestep, meaning that the local asymptotic states of flavor conversion are instantly achieved, the conversion does not achieve the equipartition value of $1/6\approx0.167$, marked by the horizontal dashed line. At relative later times, the FFC region interior to the shock shrinks and flavor conversion occurs mostly at radii of hundreds to thousands of kilometers. At such larger radii, the neutrino fractions gradually approach the equipartition value $1/6\approx0.167$. Note that even if the fraction profiles approach the equipartition value, they should never reach flavor equipartition ($\nu_e=\bar{\nu}_e=\nu_x=\bar{\nu}_x$), since that would violate the conservation of total ELN-XLN.

Instead of flavor equipartition, the ``quasi flavor equipartition'' assumption, motivated by quantum many-body calculations \citep{martin2023} and used in a phenomenological treatment of neutrino flavor conversions in binary neutron star merger simulations \citep{qiu2025,qiu2025b}, serves as a better approximation to the final flavor states of FFC conversion. This quasi flavor equipartition has four conditions\footnote{In \citet{qiu2025}, the first three conditions are expressed in a different, but equivalent, way: total neutrino number conservation, electron lepton number conservation, and heavy lepton number conservation.}:
\begin{itemize}
    \item Neutrino number conservation $N^{\rm FFC}_{\nu_e}+2N^{\rm FFC}_{\nu_x}=N_{\nu_e}+2N_{\nu_x}=N$ 
    \item Anti-neutrino number conservation $N^{\rm FFC}_{\bar{\nu}_e}+2N^{\rm FFC}_{\bar{\nu}_x}=N_{\bar{\nu}_e}+2N_{\bar{\nu}_x}=\bar{N}$
    \item Total ELN-XLN conservation $N^{\rm FFC}_{\nu_e}-N^{\rm FFC}_{\bar{\nu}_e}-2N^{\rm FFC}_{\nu_x}+2N^{\rm FFC}_{\bar{\nu}_x}=N_{\nu_e}-N_{\bar{\nu}_e}-2N_{\nu_x}+2N_{\bar{\nu}_x}=N_{\rm ELN}$ 
    \item Assumption: $\sqrt{N^{\rm FFC}_{\nu_e}N^{\rm FFC}_{\bar{\nu}_e}}=\sqrt{N^{\rm FFC}_{\nu_x}N^{\rm FFC}_{\bar{\nu}_x}}$, i.e., flavor equipartition in the sense of geometric mean,
\end{itemize}
where $N$ is the neutrino number and the superscript ``FFC'' means it's the value after the operation of the FFC. The final states can then be solved as:
\equ{
&N^{\rm FFC}_{\nu_e}=\frac{1}{12}\left(N-5\bar{N}+3N_{\rm ELN}+\sqrt{25N^2+(5\bar{N}-3N_{\rm ELN})^2+2N(7\bar{N}+15N_{\rm ELN})}\right)\\
&N^{\rm FFC}_{\bar{\nu}_e}=\frac{1}{12}\left(-5N+\bar{N}-3N_{\rm ELN}+\sqrt{25N^2+(5\bar{N}-3N_{\rm ELN})^2+2N(7\bar{N}+15N_{\rm ELN})}\right)\\
&N^{\rm FFC}_{\nu_x}=\frac{1}{24}\left(11N+5\bar{N}-3N_{\rm ELN}-\sqrt{25N^2+(5\bar{N}-3N_{\rm ELN})^2+2N(7\bar{N}+15N_{\rm ELN})}\right)\\
&N^{\rm FFC}_{\bar{\nu}_x}=\frac{1}{24}\left(5N+11\bar{N}+3N_{\rm ELN}-\sqrt{25N^2+(5\bar{N}-3N_{\rm ELN})^2+2N(7\bar{N}+15N_{\rm ELN})}\right)\\
}

Three of the four conditions used in the quasi flavor equipartition formalism are conservation laws in FFC, so the performance of this method solely relies on the quality of the extra assumption $\sqrt{N^{\rm FFC}_{\nu_e}N^{\rm FFC}_{\bar{\nu}_e}}=\sqrt{N^{\rm FFC}_{\nu_x}N^{\rm FFC}_{\bar{\nu}_x}}$. Figure \ref{fig:asymptotic} shows the neutrino number density fraction profiles of $\sqrt{\nu_e\bar{\nu}_e}$ and $\sqrt{\nu_x\bar{\nu}_x}$ at two representative time snapshots. In terms of number density fraction, the quasi equipartition assumption means that the fractions of $\sqrt{\nu_e\bar{\nu}_e}$ and $\sqrt{\nu_x\bar{\nu}_x}$ are both equal to $1/6\approx0.167$. At early times ($\sim$50 ms post-bounce), although the FFC has already resulted in a $\sim20$\% flavor conversion, the final flavor state is far from quasi equipartition. However, quasi-equipartition becomes a very good approximation at later times ($\sim$300 ms), since the number fractions of $\sqrt{\nu_e\bar{\nu}_e}$ and $\sqrt{\nu_x\bar{\nu}_x}$ both quickly converge to the equipartition value.

The above formulae provide the FFC altered neutrino number luminosities, and to get the energy luminosities we need the average neutrino energies. One straightforward assumption is that $\langle E^{\rm FFC}_{\nu_e}\rangle=\langle E^{\rm FFC}_{\nu_x}\rangle=(N_{\nu_e}\langle E_{\nu_e}\rangle+2N_{\nu_x}\langle E_{\nu_x}\rangle)/(N_{\nu_e}+2N_{\nu_x})$ and $\langle E^{\rm FFC}_{\bar{\nu}_e}\rangle=\langle E^{\rm FFC}_{\bar{\nu}_x}\rangle=(N_{\bar{\nu}_e}\langle E_{\bar{\nu}_e}\rangle+2N_{\bar{\nu}_x}\langle E_{\bar{\nu}_x}\rangle)/(N_{\bar{\nu}_e}+2N_{\bar{\nu}_x})$, since they are mixed by flavor conversion. This behavior has been confirmed by Figure \ref{fig:spectra}.

\begin{figure*}
    \centering
    \includegraphics[width=0.48\textwidth]{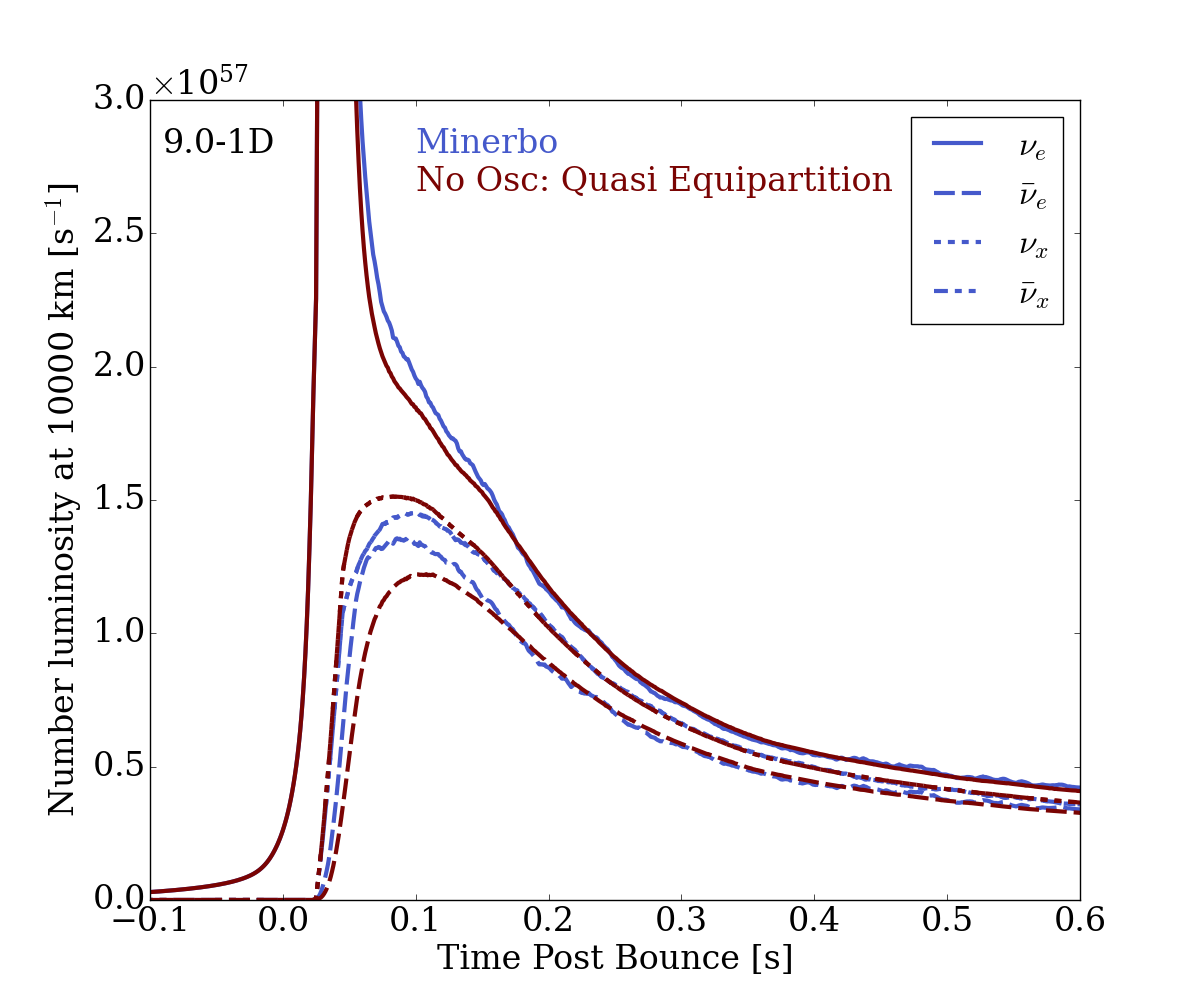}
    \includegraphics[width=0.48\textwidth]{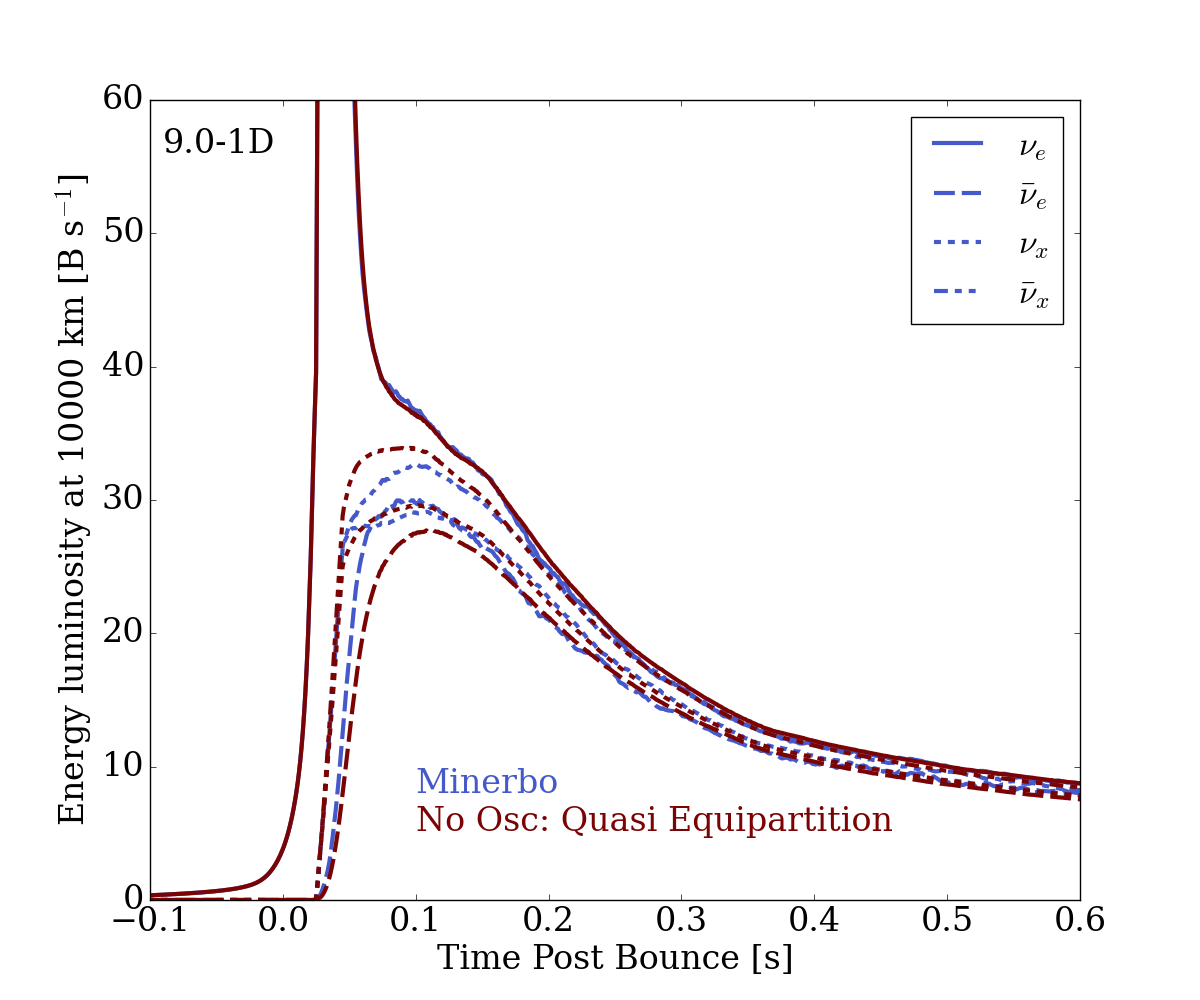}
    \includegraphics[width=0.48\textwidth]{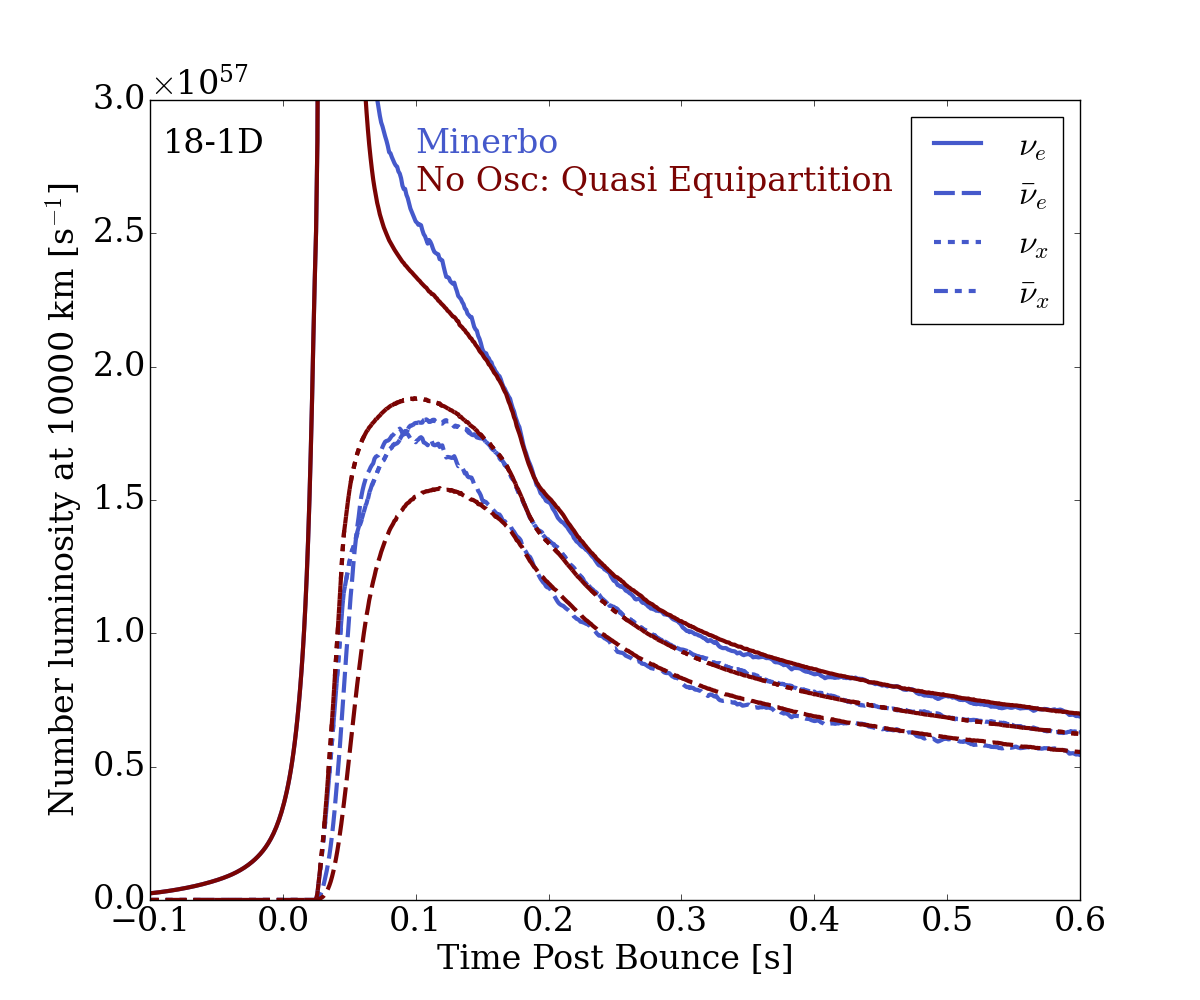}
    \includegraphics[width=0.48\textwidth]{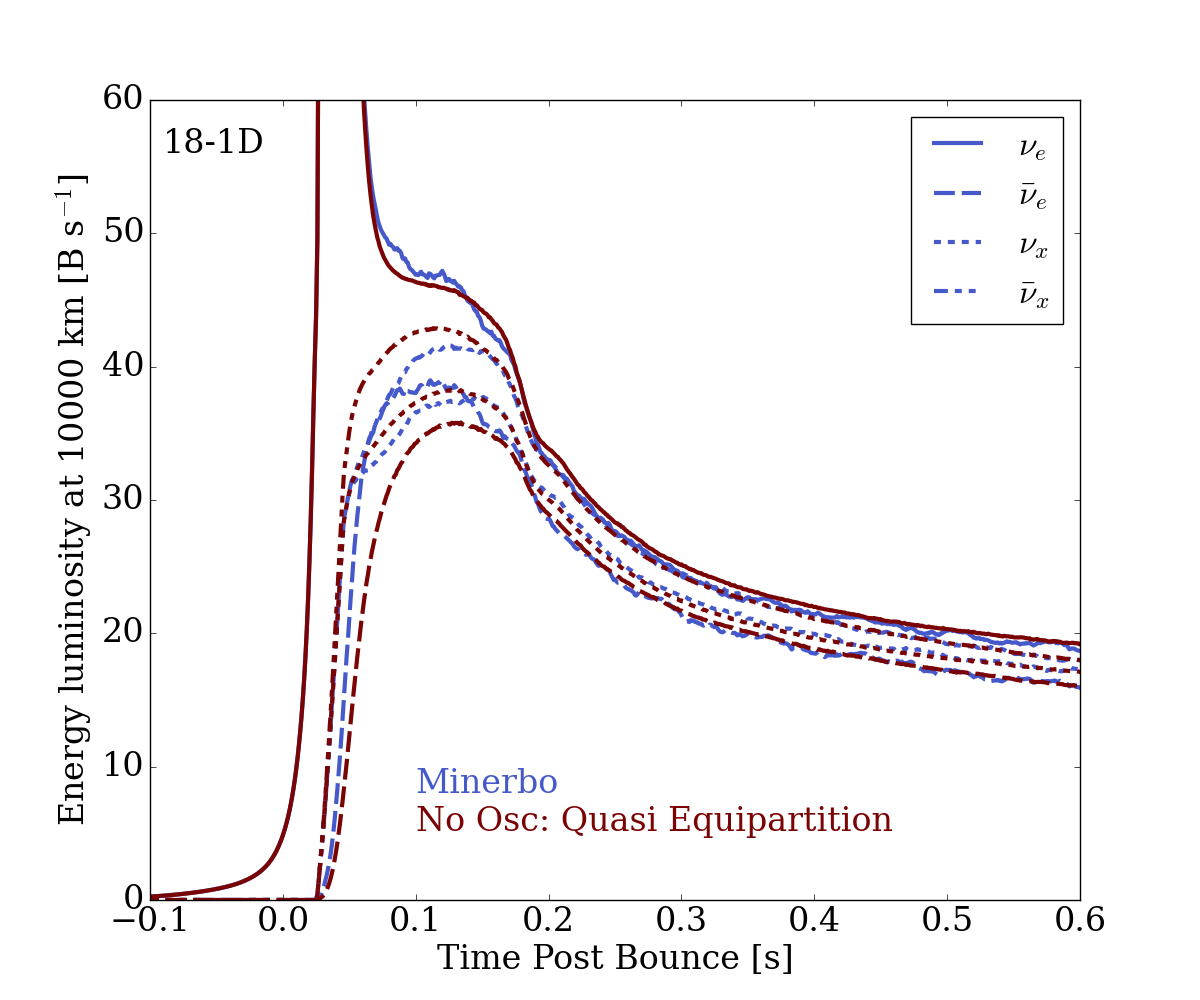}
    \caption{The neutrino number luminosity and energy luminosity evolution for the 1D FFC models compared with the post-processed results from the 1D no oscillation model calculated using the quasi equipartition approximation. Since different FFC implementations show very similar behavior, we choose to show only the Minerbo closure results on this plot because they use the same closure as the no-oscillation models. It can be seen that both the neutrino number luminosities and the neutrino energy luminosities are well-approximated by the quasi-equipartition method, especially at late times. The relative errors at late times are less than 2\%, while slightly larger deviations are found at early times (within about 200 ms post-bounce).}
    \label{fig:quasi}
\end{figure*}

Figure \ref{fig:quasi} shows the neutrino number and energy luminosities calculated in the 1D models with our Box3D-BGK scheme and the post-processed results calculated from the no-oscillation 1D model using the quasi-equipartition formulae. Since models with FFC all behave similarly, we show only the Minerbo models here because they use the same closure as the no-oscillation models. Since it can be derived that the quasi-equipartition method conserves $N_{\nu_x}-N_{\bar{\nu}_x}$, the resultant $\nu_x$ and $\bar{\nu}_x$ number luminosities remain identical and, therefore, overlap on the plot. Despite the fact that this phenomenological method cannot capture the number differences between $\nu_x$ and $\bar{\nu}_x$, it can be seen that both the neutrino number luminosities and the neutrino energy luminosities are well-approximated, especially at late times. The relative errors at late times are less than 2\%, while slightly larger deviations are found at early times (near $\sim$200 ms post-bounce).

With this quasi-equipartition approximation, one can estimate the FFC-altered neutrino properties by post-processing the neutrino signals extracted from no-oscillation CCSN simulations. This phenomenological method provides a simple way to include the effects of FFC on neutrino signals without implementing a complex and expensive FFC scheme and re-doing the simulations.

\subsection{Two-Dimensional Models}
The 1D models capture the major behaviors of the FFC, but they are mute concerning several important features of modern CCSNe models. First, 1D models typically don't explode and the FFC effects after the launch of an explosion would therefore remain unknown. Second, multi-dimensional effects, such as convection and turbulence, may change the structure of the neutrino radiation field, and this might have an impact on the occurence of the FFC. To tackle such questions, we now turn to 2D simulations for the same progenitors discussed above. 

\begin{figure*}
    \centering
    \includegraphics[width=0.48\textwidth]{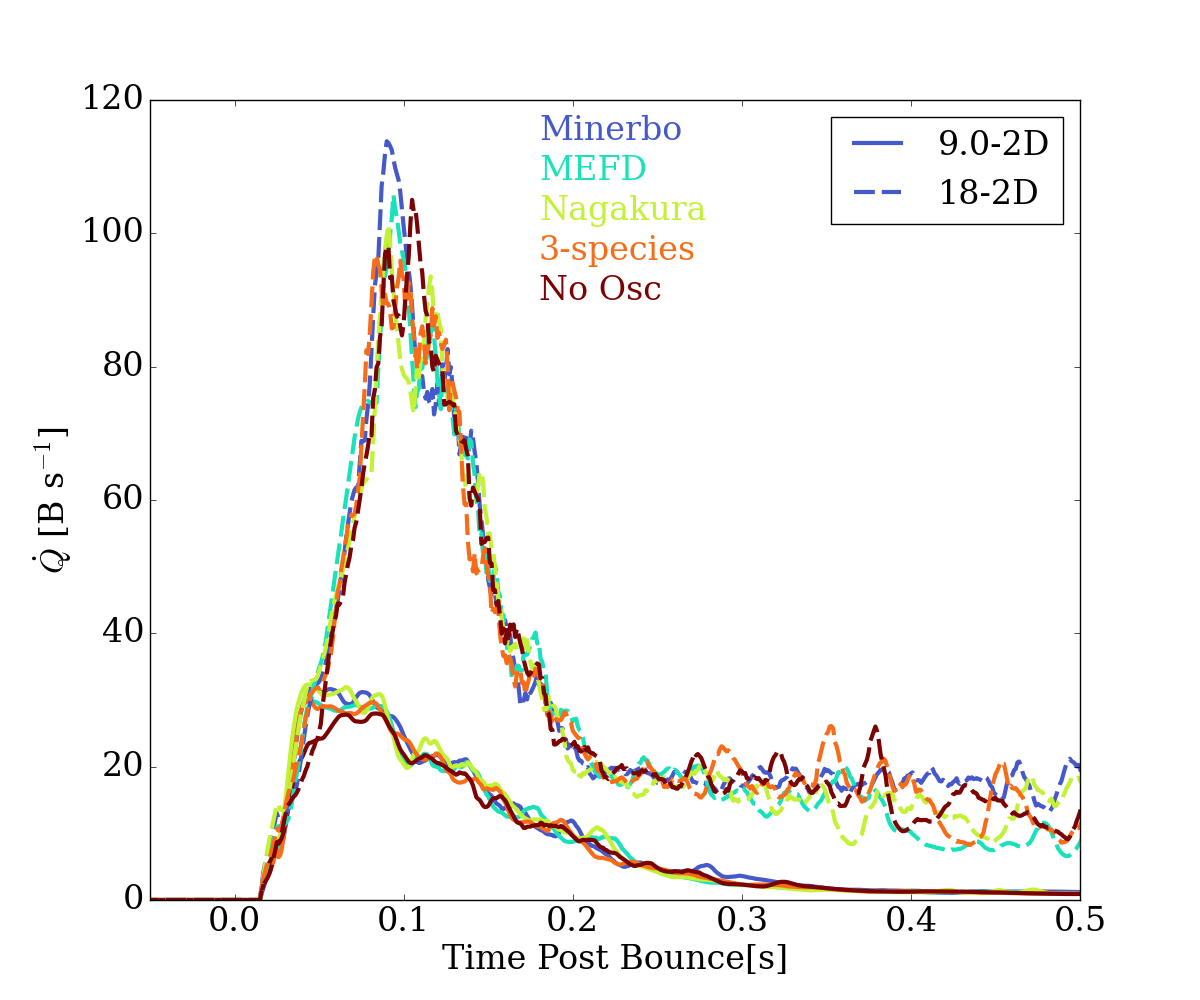}
    \includegraphics[width=0.48\textwidth]{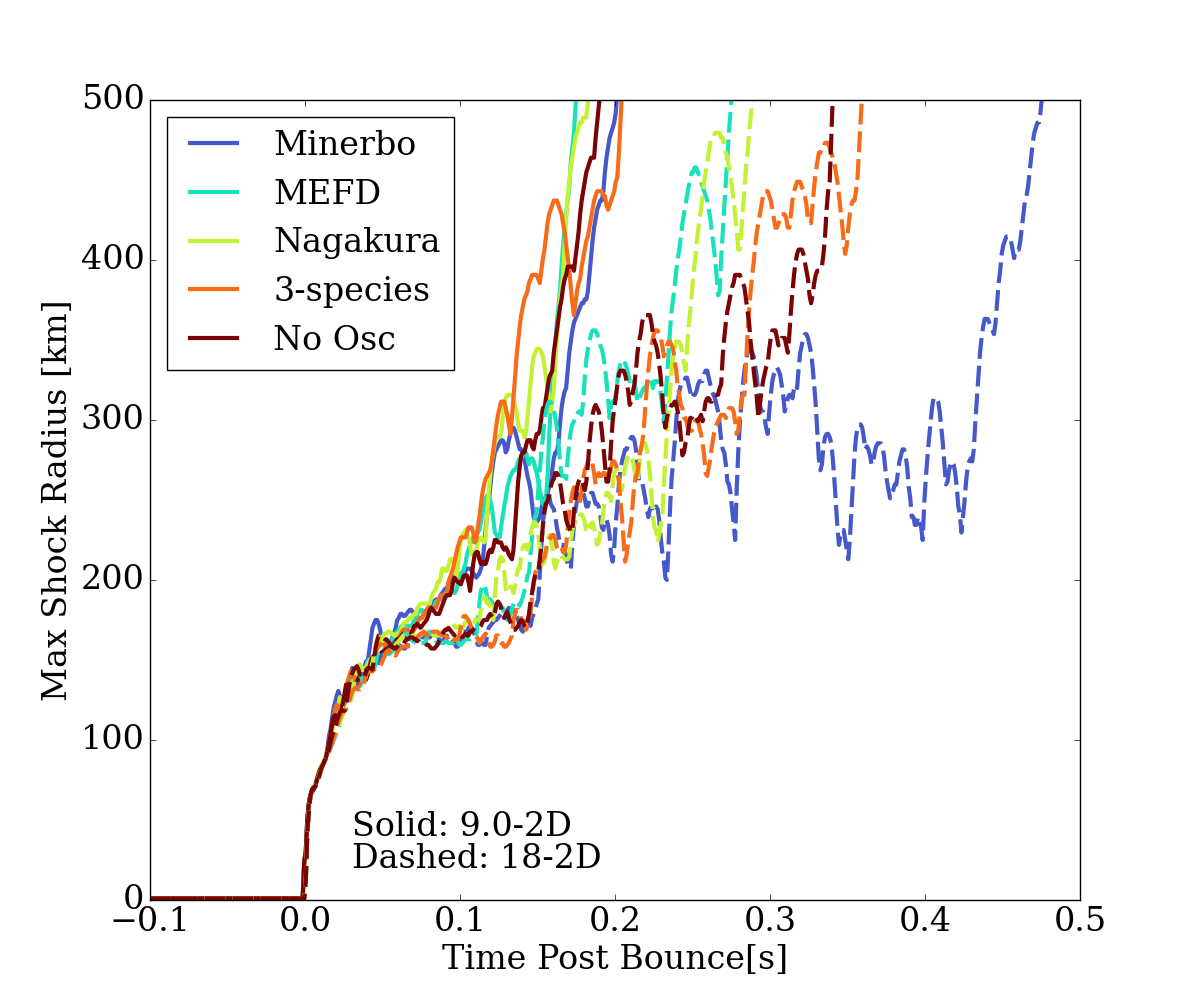}
    \caption{Left: Net neutrino heating rate (in unit of $10^{51}$ erg $s^{-1}$) in the gain region behind the shock for the 2D models. Right: the maximum shock radius as a function of time. The 9 M$_\odot$ model is shown using solid lines, while the 18 M$_\odot$ model is marked using dashed lines. Different colors indicate the different treatments of neutrino fast flavor conversion. Although multi-dimensional effects, such as turbulence and convection, make the behavior more stochastic, the general trend is the same as found in 1D (Figure \ref{fig:heating}): Models that include FFC (``Minerbo'', ``MEFD'', ``Nagakura'', and ``3-species'') all behave very similar to each other, and they show slightly higher heating rates at early times in the 9 M$_\odot$ model, compared to the no oscillation model (``No osc''). However, the overall effects of FFC on heating rates and shock evolution are weak. This confirms the findings in \citet{wang2025}.}
    \label{fig:heating-2d}
\end{figure*}

Figure \ref{fig:heating-2d} shows the net neutrino heating rates and maximum shock radius evolution for the 2D models. Despite of the larger fluctuations, the major behaviors follow what we saw in 1D models. Models with FFC effects behave very similarly. After a short delay post-bounce, the net heating rates in the 9 M$_\odot$ model are enhanced by about 10\% for a few tens of milliseconds and then settle to that witnessed without the FFC effects. In the 18 M$_\odot$, no significant FFC effects on the net heating rates are seen. 

The FFC effects on shock radius evolution are also minor. For the 9 M$_\odot$ models, the FFC effects result in slightly earlier explosion times (by about a few milliseconds) than the no oscillation model, while for the 18 M$_\odot$ model there is no significant explosion time difference. This is consistent with the behavior of the net neutrino heating rates.

\begin{figure*}
    \centering
\includegraphics[width=0.48\textwidth]{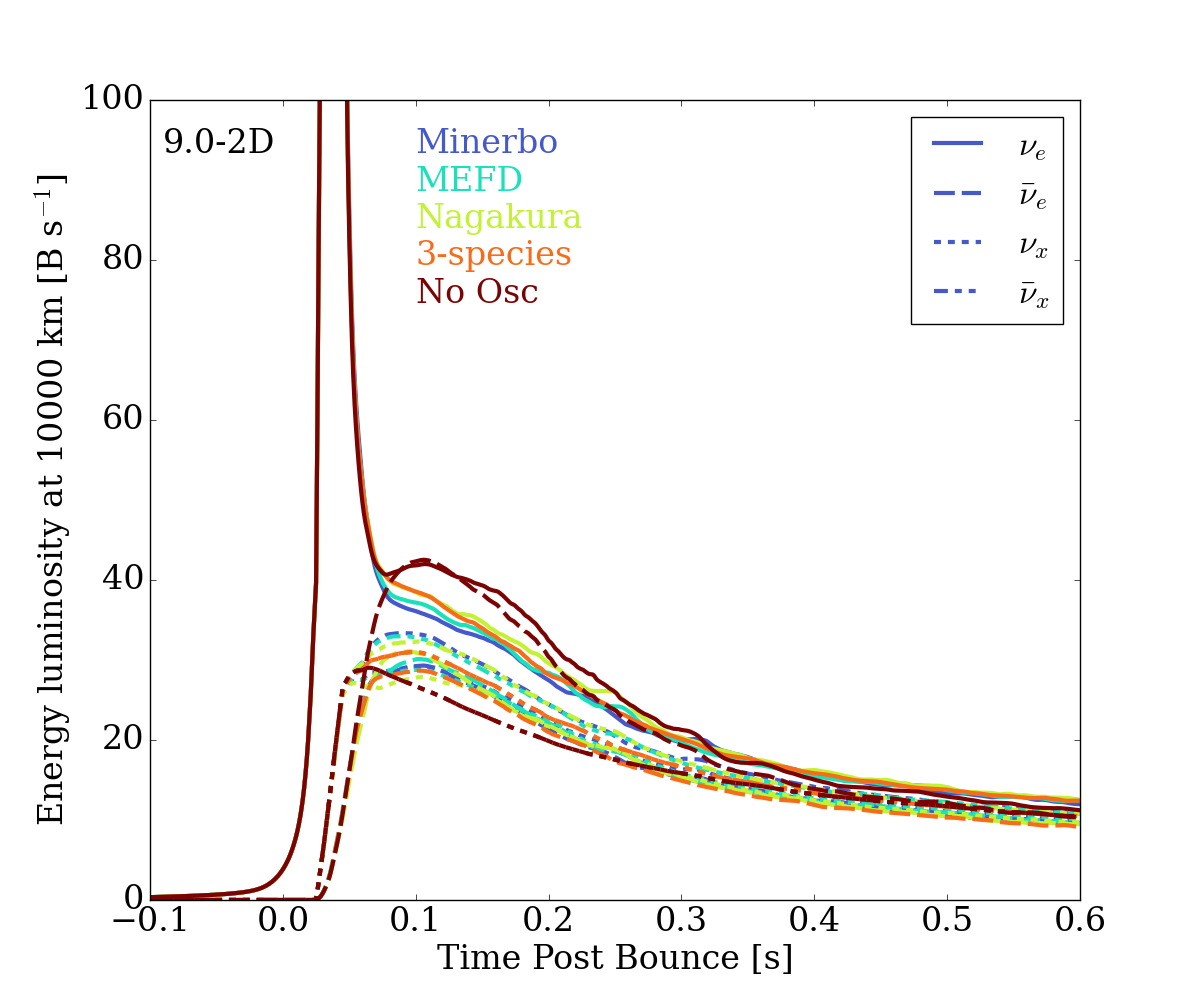}
    \includegraphics[width=0.48\textwidth]{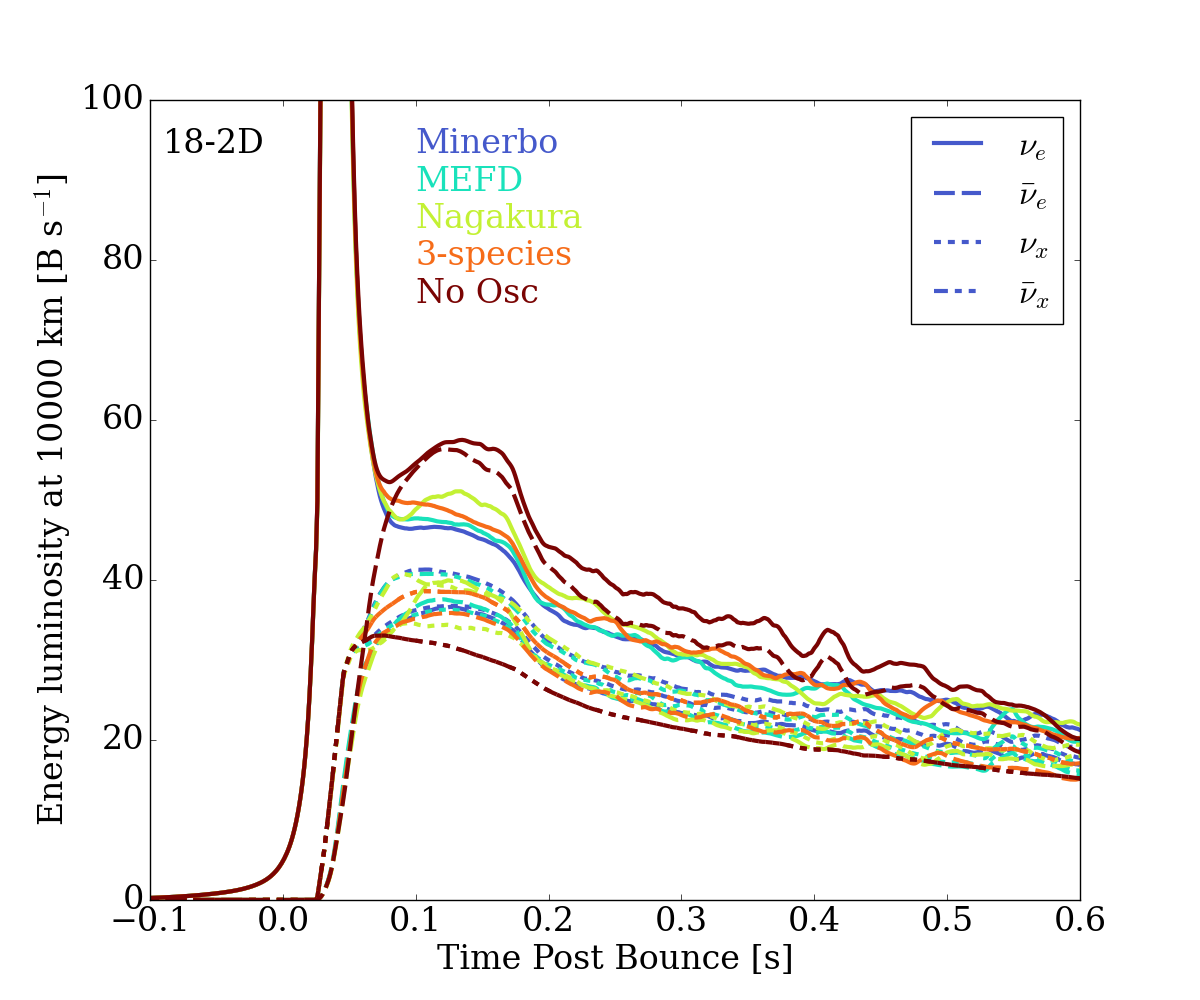}
    \caption{Same as Figure \ref{fig:luminosities}, but for 2D models. Compared with the 1D models, the 2D luminosity curves show more intrinsic fluctuations due to multi-dimensional effects such as turbulence and convection. Such fluctuations are at a similar level to the differences between the results for various angular reconstruction methods and FFC schemes, which indicates that the uncertainties introduced by the FFC schemes are not very significant. }
    \label{fig:luminosities-2d}
\end{figure*}

Figure \ref{fig:luminosities-2d} shows the time-dependent neutrino energy luminosities for the 2D models. At early times, about 20\% of the electron and anti-electron neutrino luminosities are converted into the $x$- and anti-$x$-type neutrino luminosities by the FFC, while this fraction decreases at later times. Such behaviors are similar to the 1D 9 M$_\odot$ model, while the 1D 18 M$\odot$ model shows larger conversion fractions at late times. This is because the successful explosion in 2D dimishes the mass accretion rate onto the proto-neutron star and changes the neutrino luminosities. We expect the non-exploding massive progenitors, which maintain stronger accretion flows, to behave more similarly to the 1D 18 M$_\odot$ model and to show higher conversion fractions at relatively later times.

\begin{figure*}
    \centering
    \includegraphics[width=0.48\textwidth]{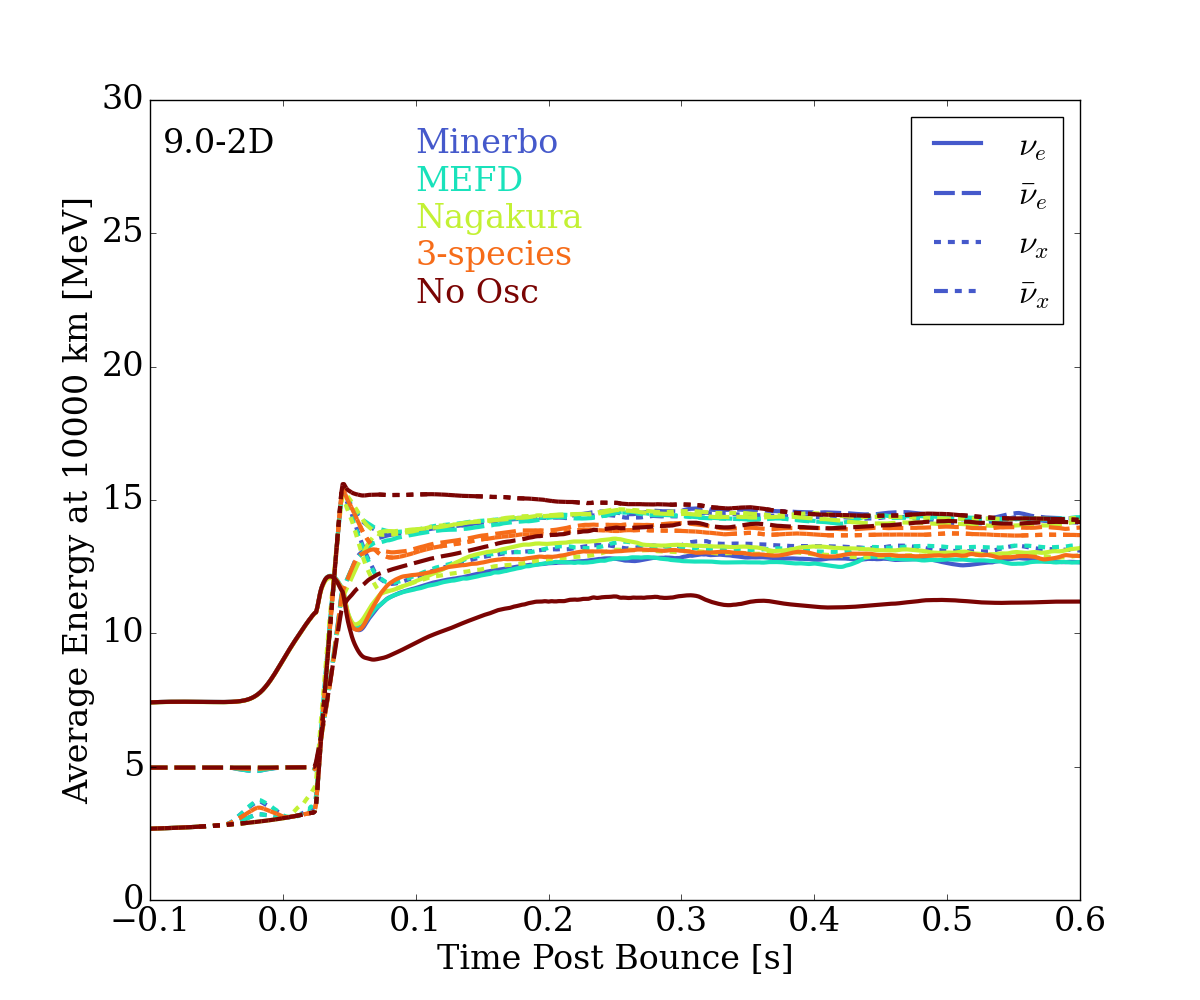}
    \includegraphics[width=0.48\textwidth]{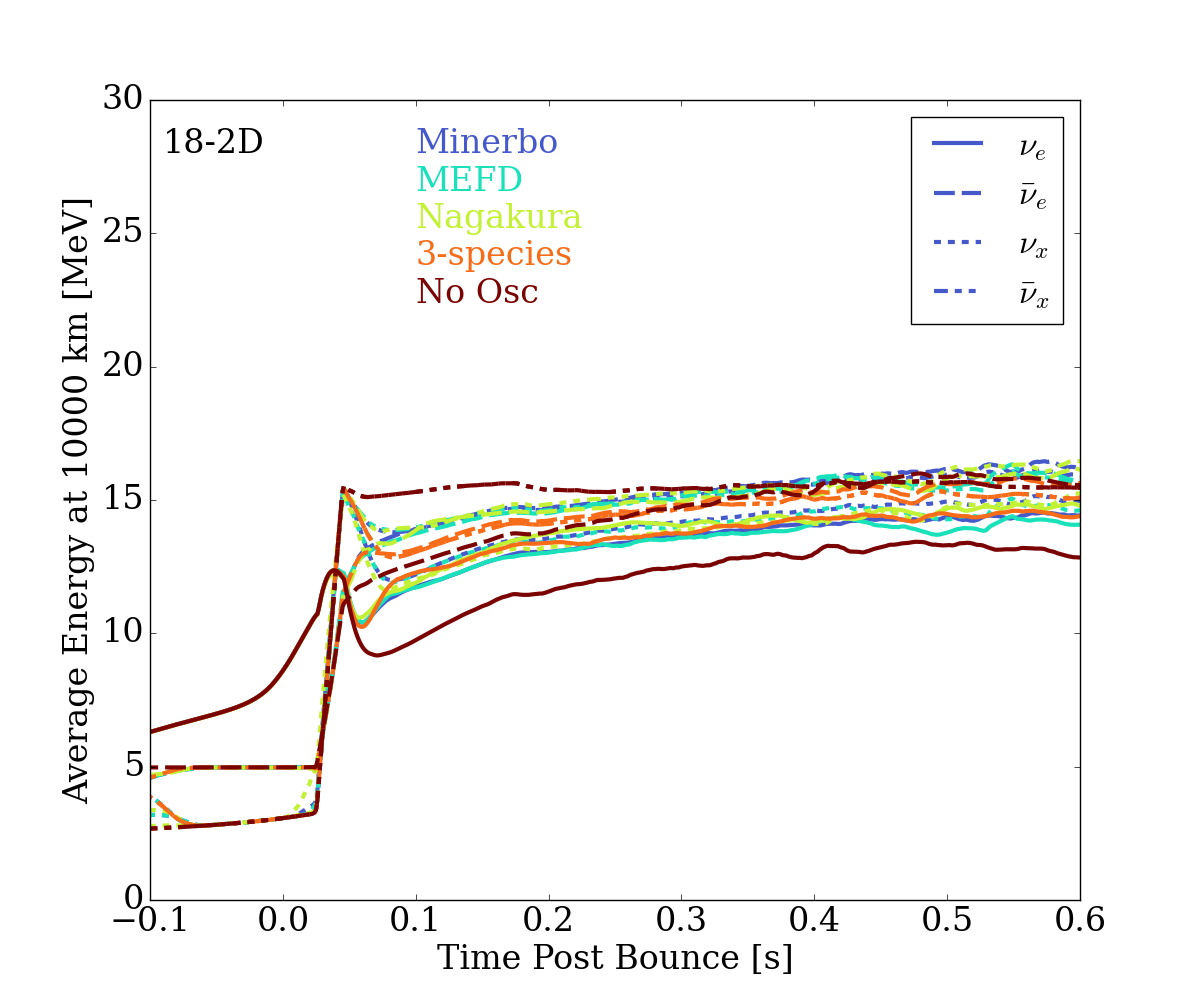}
    \caption{Same as Figure \ref{fig:spectra}, but for the 2D models. The same behavior is seen here: When the FFC is operative, the $\nu_{x}$ and $\bar{\nu}_{x}$ neutrino spectra are softened, while both the $\nu_e$ and $\bar{\nu}_e$ neutrino spectra harden slightly. The differences between various angular reconstruction methods are minor. With the FFC, the average energies of the four neutrino types show smaller, but non-vanishing, differences.  Neutrinos ($\nu_e$ and $\nu_x$) have similar average energies, while anti-neutrinos ($\bar{\nu}_e$ and $\bar{\nu}_x$) have similar average energies, but at higher values. } 
    \label{fig:spectra-2d}
\end{figure*}

In 2D and enabling the FFC, the deviations between the energy luminosities of the various neutrino types are larger than in 1D. However, it is clear that such deviations decrease over time. The luminosity order $L_{\nu_e}>L_{\bar{\nu}_x}>L_{\nu_x}>L_{\bar{\nu}_e}$ still holds for most of the time.

Figure \ref{fig:spectra-2d} shows the evolution of the average energies of different neutrino types for the 2D models. All FFC models show similar behaviors regardless of the exact implementation of the FFC scheme. When the FFC is operative, the $\nu_{x}$ and $\bar{\nu}_{x}$ neutrino spectra are softened, while both the $\nu_e$ and $\bar{\nu}_e$ neutrino spectra harden slightly.  With FFC, the average energies of the four neutrino types show smaller, but non-vanishing, differences.  Neutrinos ($\nu_e$ and $\nu_x$) have similar average energies, while anti-neutrinos ($\bar{\nu}_e$ and $\bar{\nu}_x$) have similar average energies, but at higher values. 

\begin{figure*}
    \centering
    \includegraphics[width=0.48\textwidth]{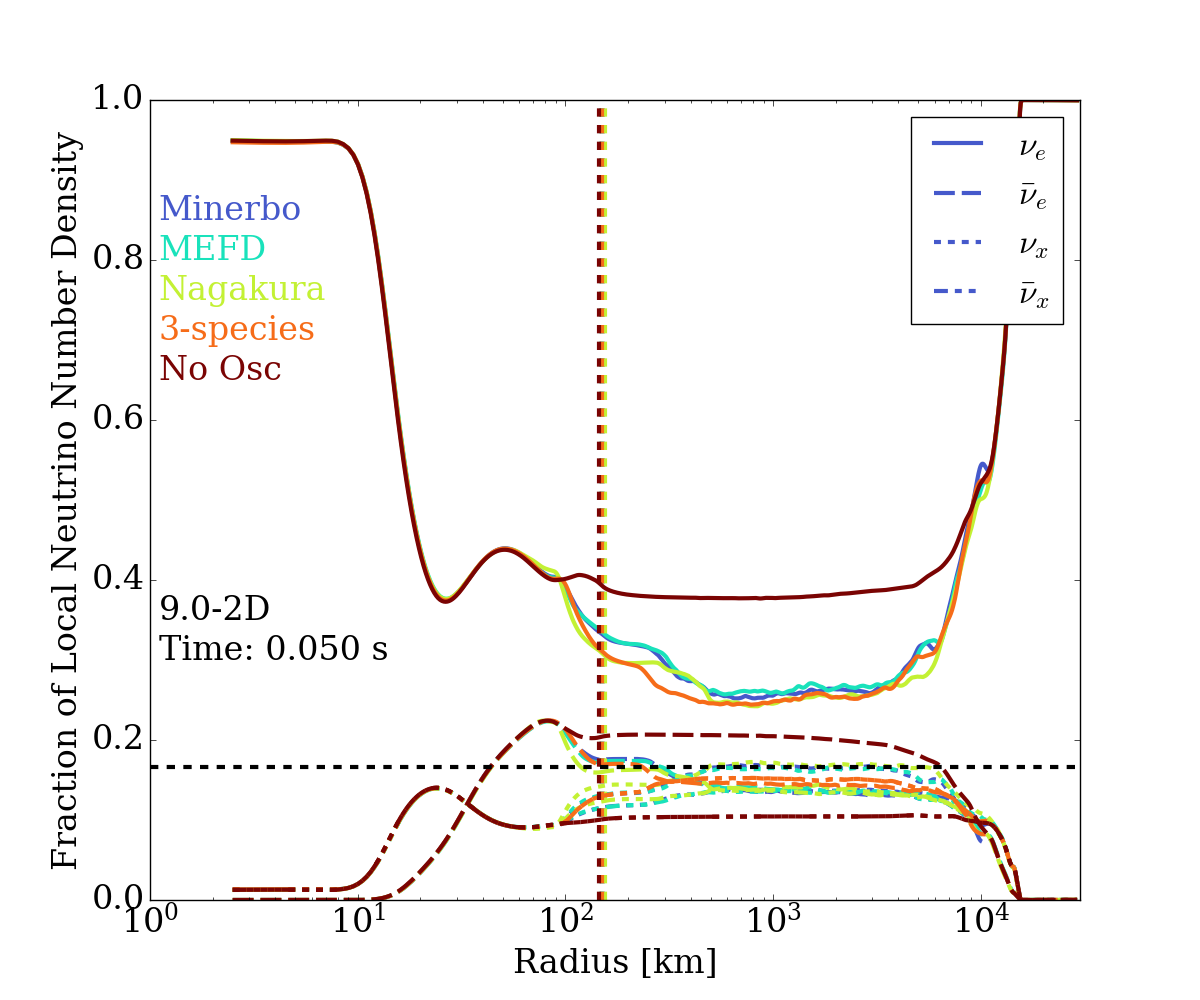}
    \includegraphics[width=0.48\textwidth]{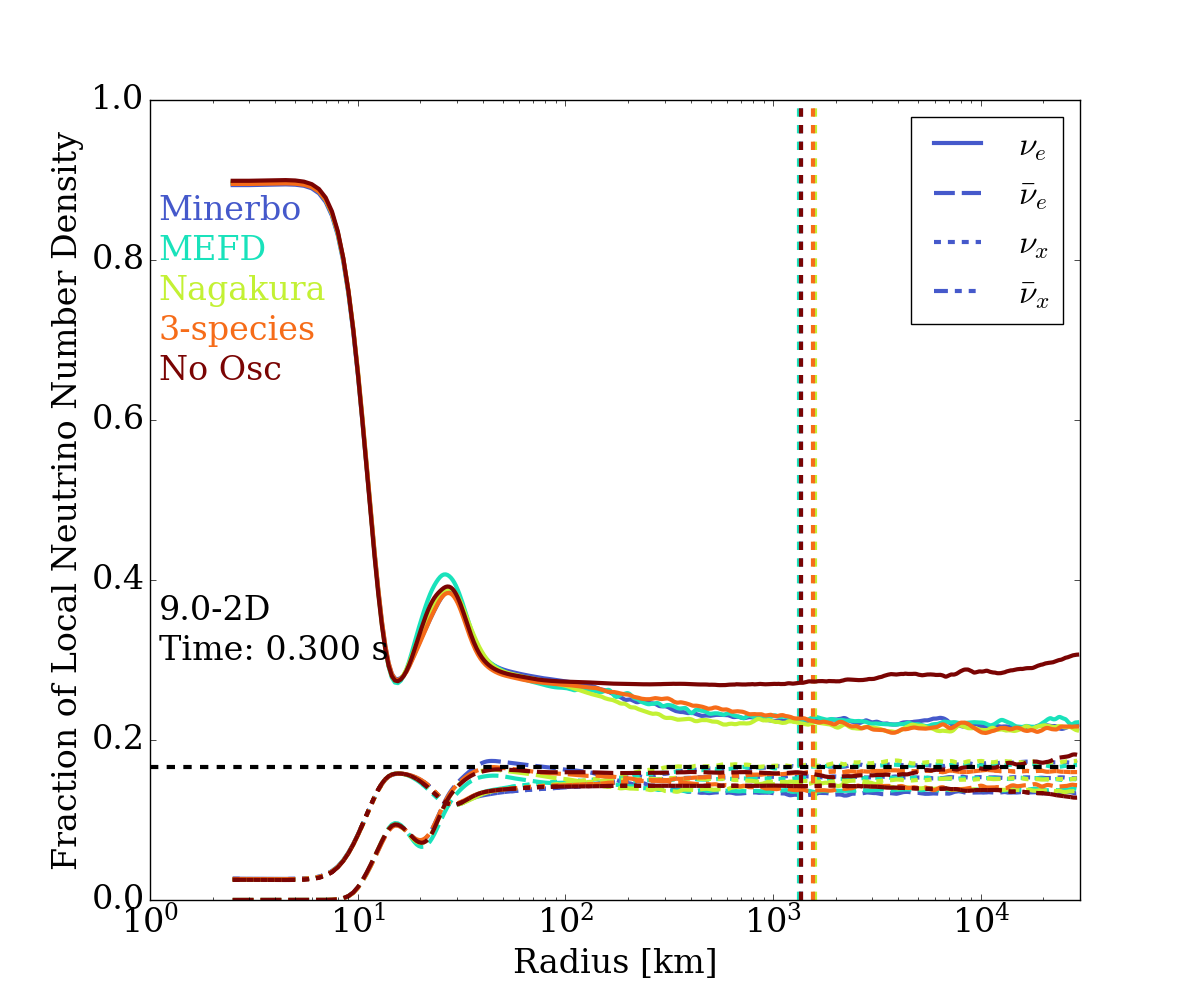}
    \includegraphics[width=0.48\textwidth]{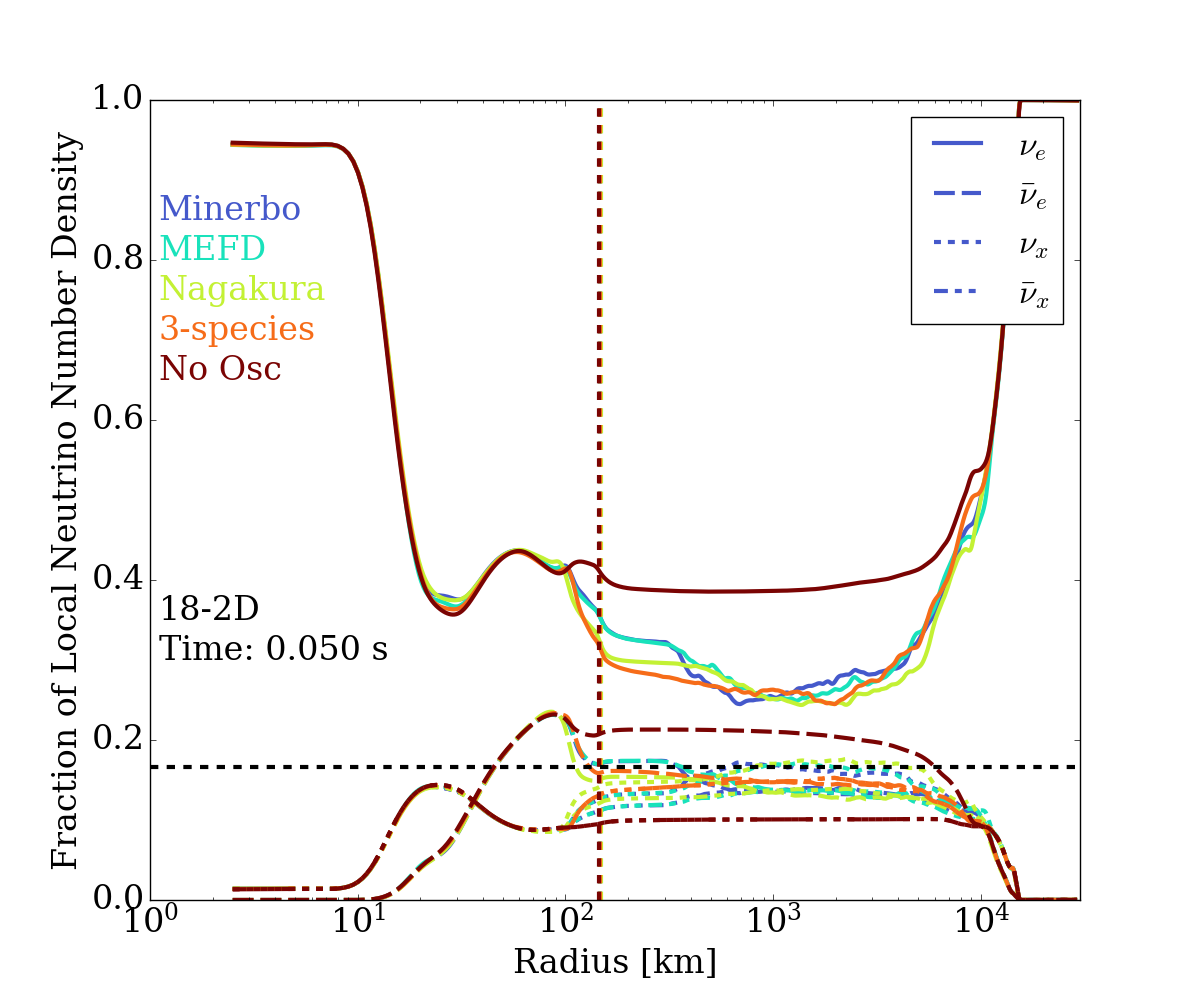}
    \includegraphics[width=0.48\textwidth]{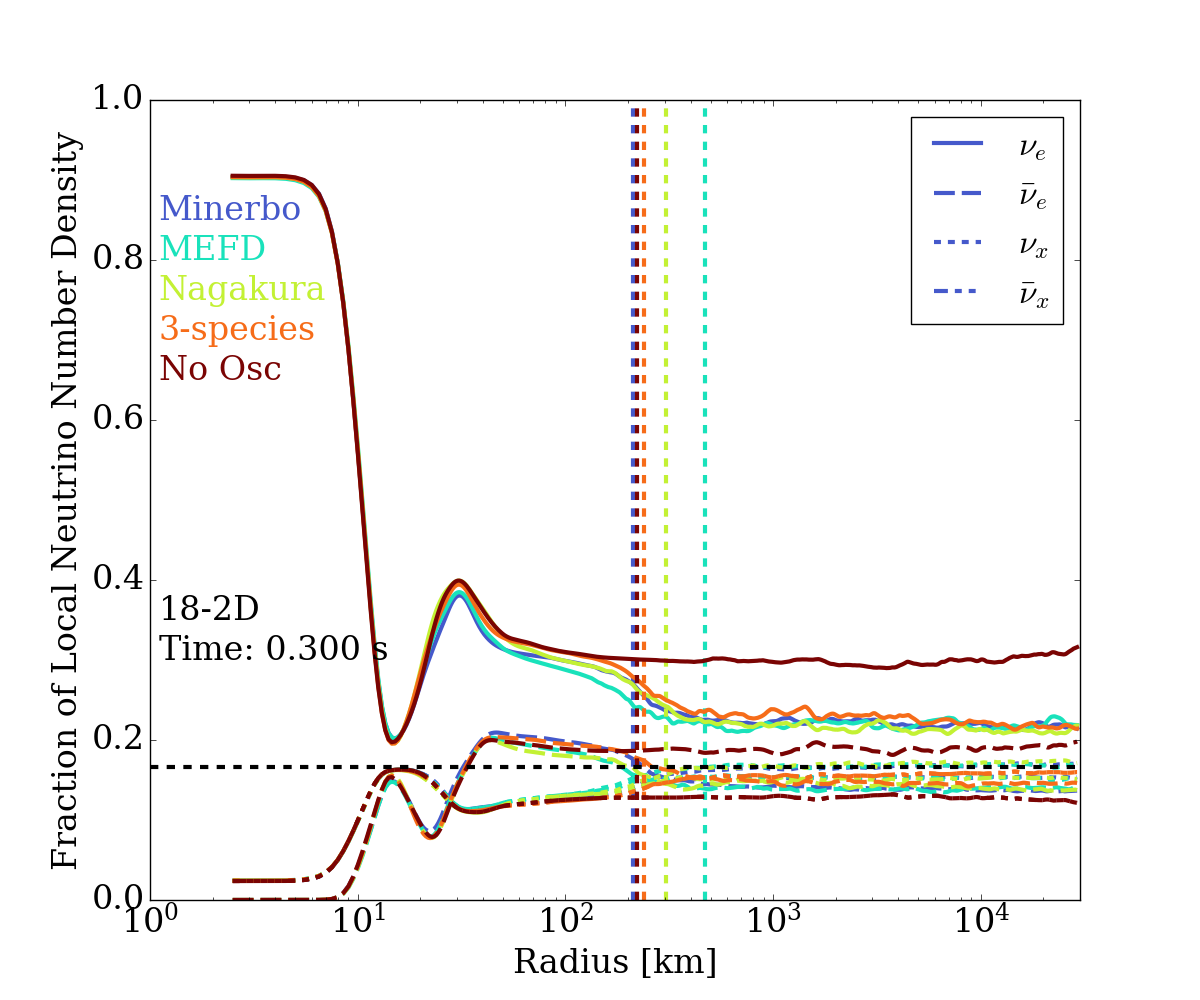}
    \caption{Same as Figure \ref{fig:fractions}, but for the 2D models. At early times ($\sim$50 ms post-bounce), the conversion happens mostly interior to the shock ($<\sim200$ km). At relatively later times ($\sim$300 ms post-bounce), flavor conversion happens at hundreds to thousands of kilometers and flavor equipartition is gradually approached at large radii. Compared with the 1D models, the explosion of the 2D models doesn't lead to significant differences.}

    \label{fig:fractions-2d}
\end{figure*}

Figure \ref{fig:fractions-2d} shows the neutrino number density fraction profiles for the 2D models at two selected snapshots. We see very similar results as in the 1D cases. Shortly after core bounce, the FFC mostly happens in the outer neutrino gain region interior to the shock. At relatively later times, the FFC region interior to the shock shrinks and flavor conversion occurs mostly at radii of hundreds to thousands of kilometers. In these metrics, and compared to what we witness in 1D models, the successful explosions in 2D models don't lead to significant differences.

\begin{figure*}
    \centering
    \includegraphics[width=0.48\textwidth]{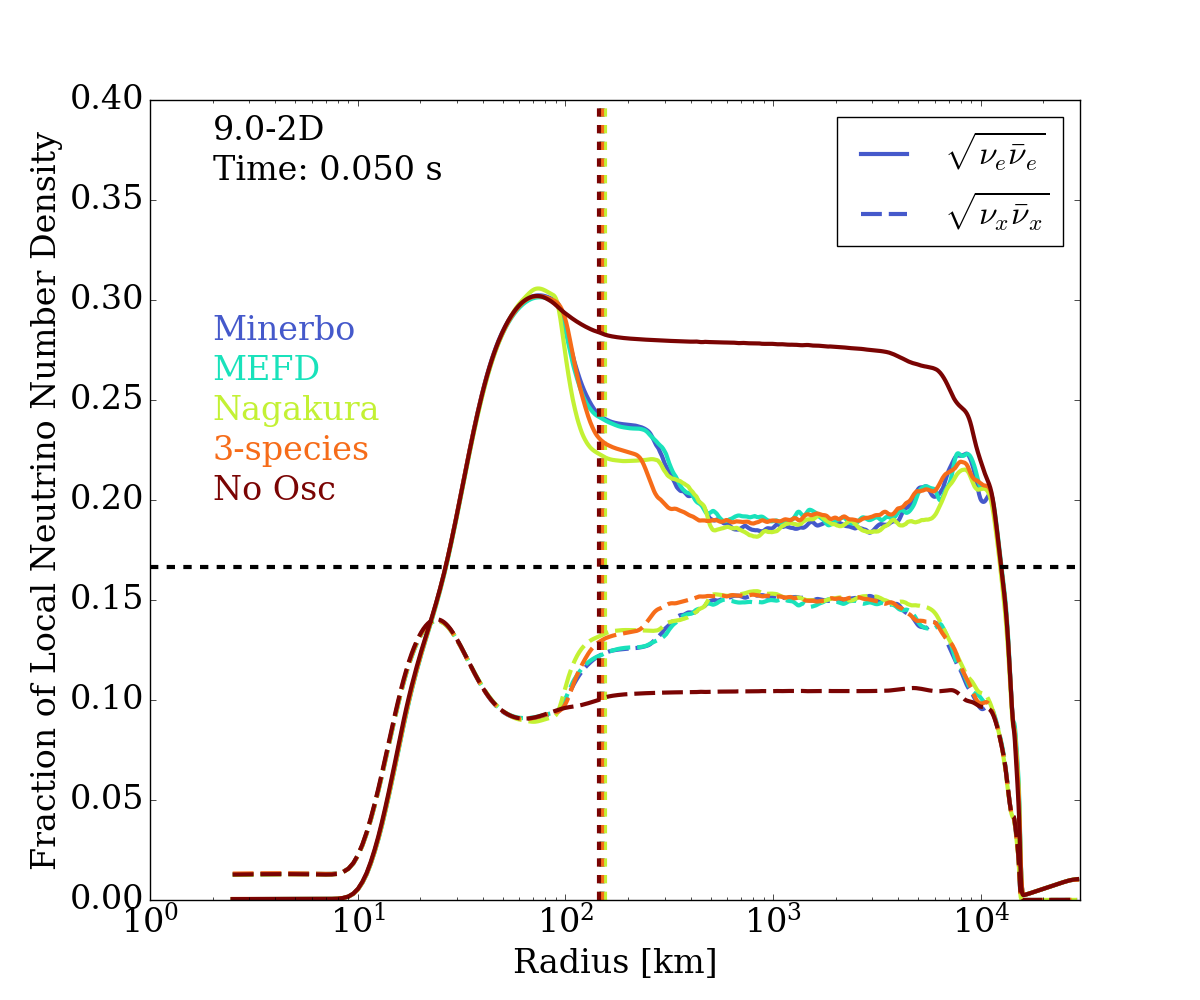}
    \includegraphics[width=0.48\textwidth]{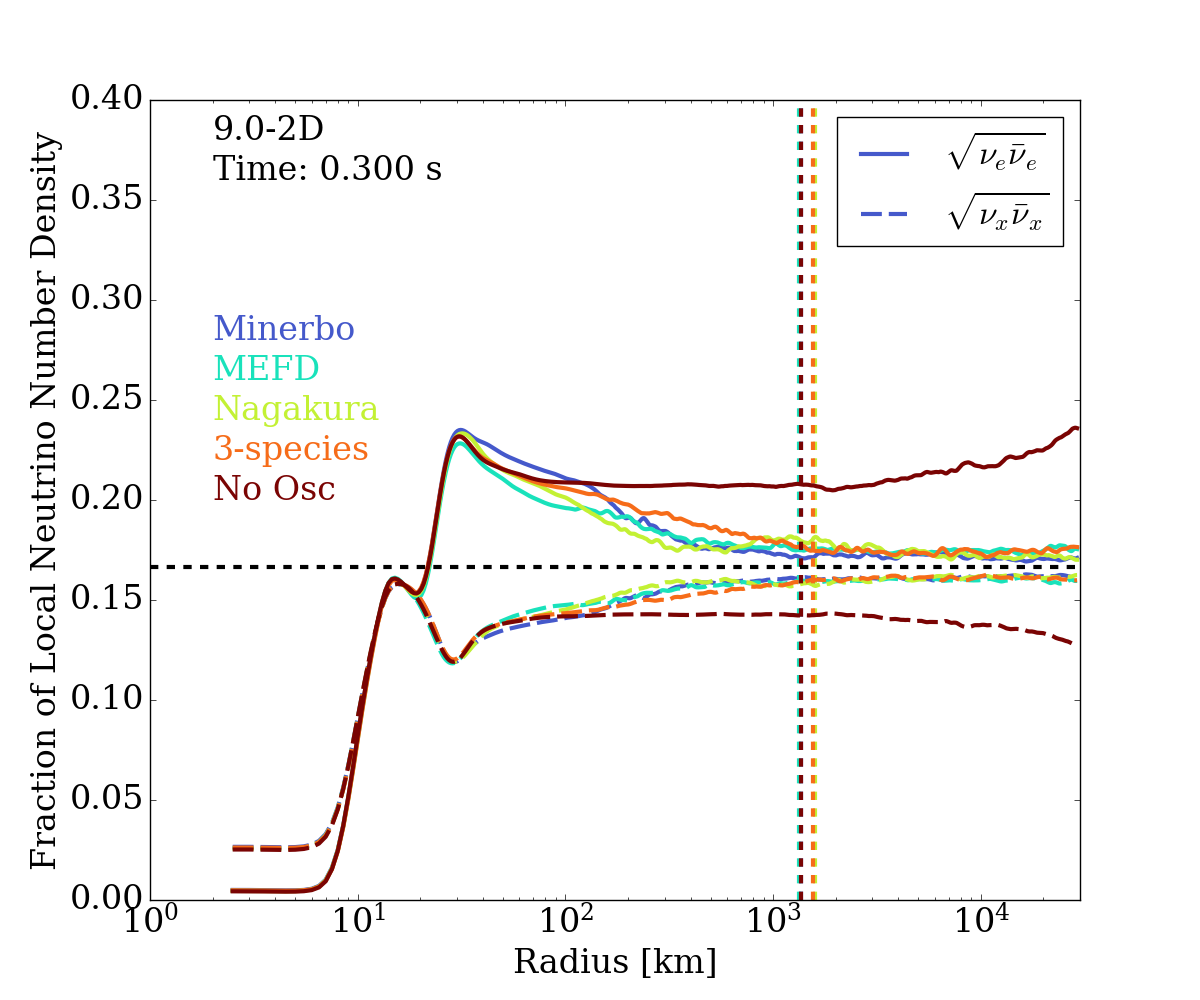}
    \includegraphics[width=0.48\textwidth]{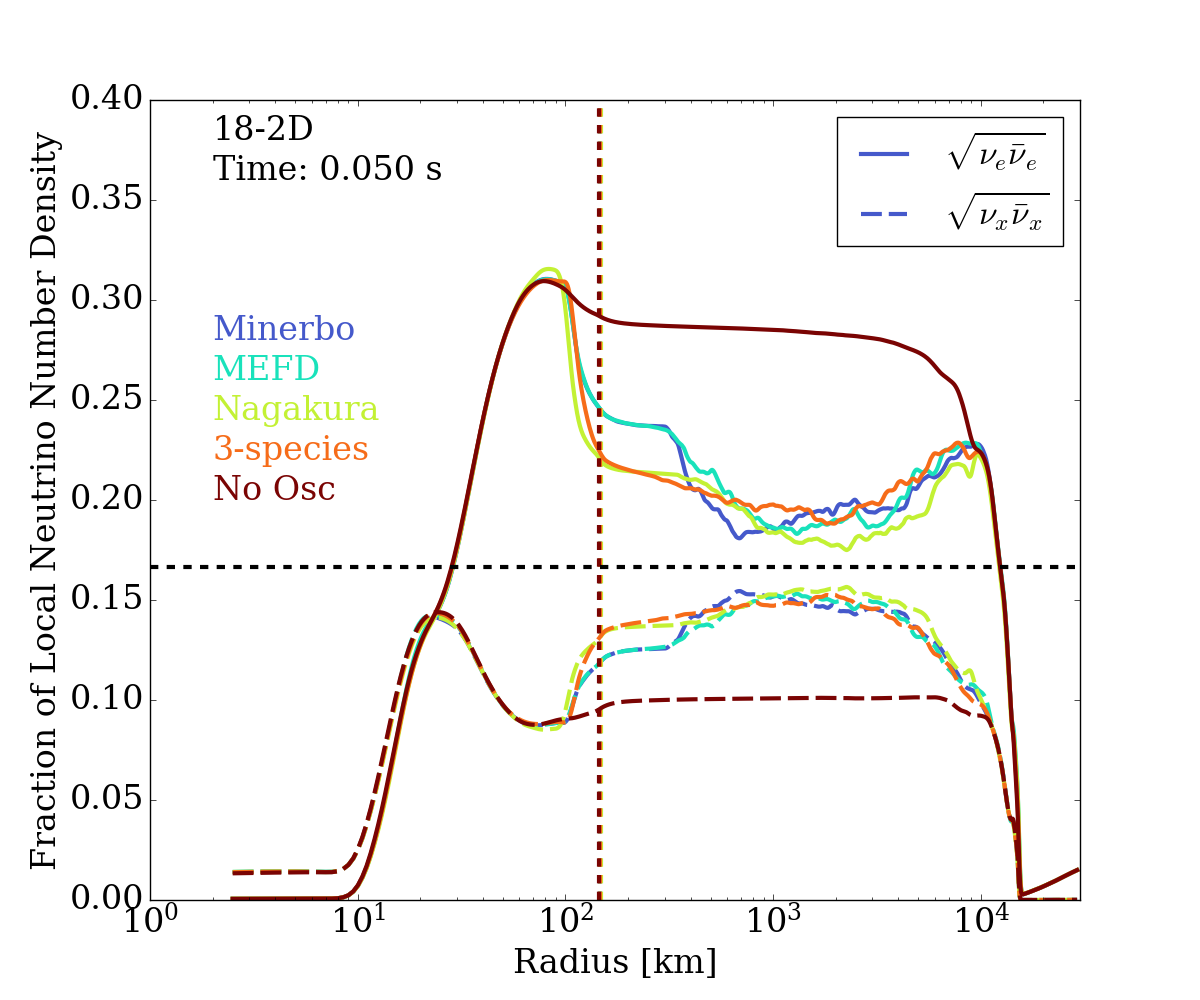}
    \includegraphics[width=0.48\textwidth]{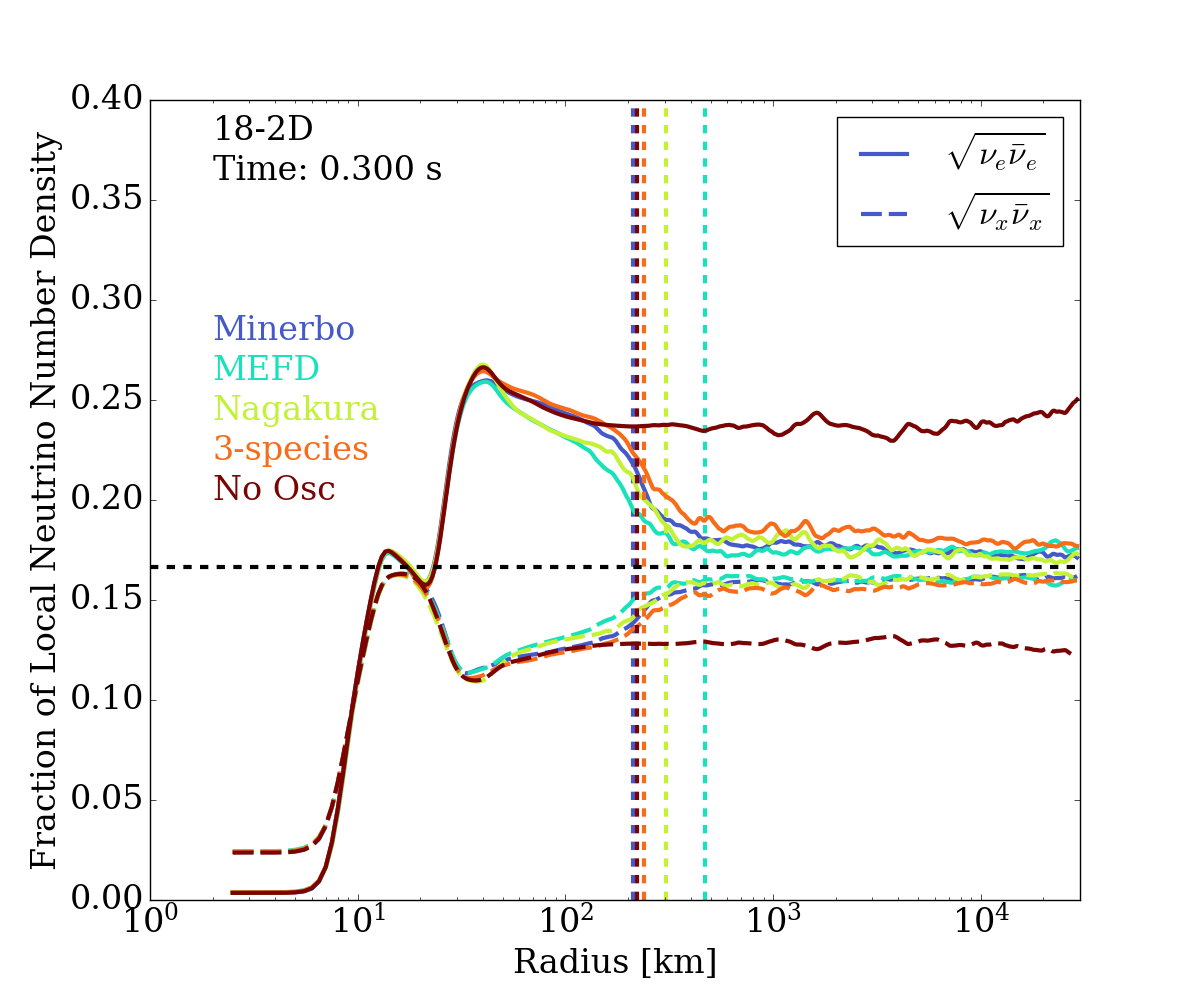}
    \caption{Same as Figure \ref{fig:asymptotic}, but for 2D models. Although at early times ($\sim$50 ms) the neutrino number density fraction doesn't satisfy $\sqrt{N^{\rm FFC}_{\nu_e}N^{\rm FFC}_{\bar{\nu}_e}}=\sqrt{N^{\rm FFC}_{\nu_x}N^{\rm FFC}_{\bar{\nu}_x}}$, the quasi-equipartition condition serves as a good approximation at later times ($\ge$300 ms post-bounce). }
    \label{fig:asymptotic-2d}
\end{figure*}

Figure \ref{fig:asymptotic-2d} shows the neutrino number density fraction profiles of $\sqrt{\nu_e\bar{\nu}_e}$ and $\sqrt{\nu_x\bar{\nu}_x}$ at two representative time snapshots for the 2D models. This behavior is similar to that of 1D models. The final flavor state is far from quasi equipartition at early times ($\sim$50 ms post-bounce), while the number fractions of $\sqrt{\nu_e\bar{\nu}_e}$ and $\sqrt{\nu_x\bar{\nu}_x}$ both approach the equipartition value ($1/6\approx0.167$) at large radii by later times ($\sim$300 ms post-bounce). This indicates that the quasi-equipartition assumption works well after a few hundreds of milliseconds post-bounce in 2D as well.

\begin{figure*}
    \centering
    \includegraphics[width=0.48\textwidth]{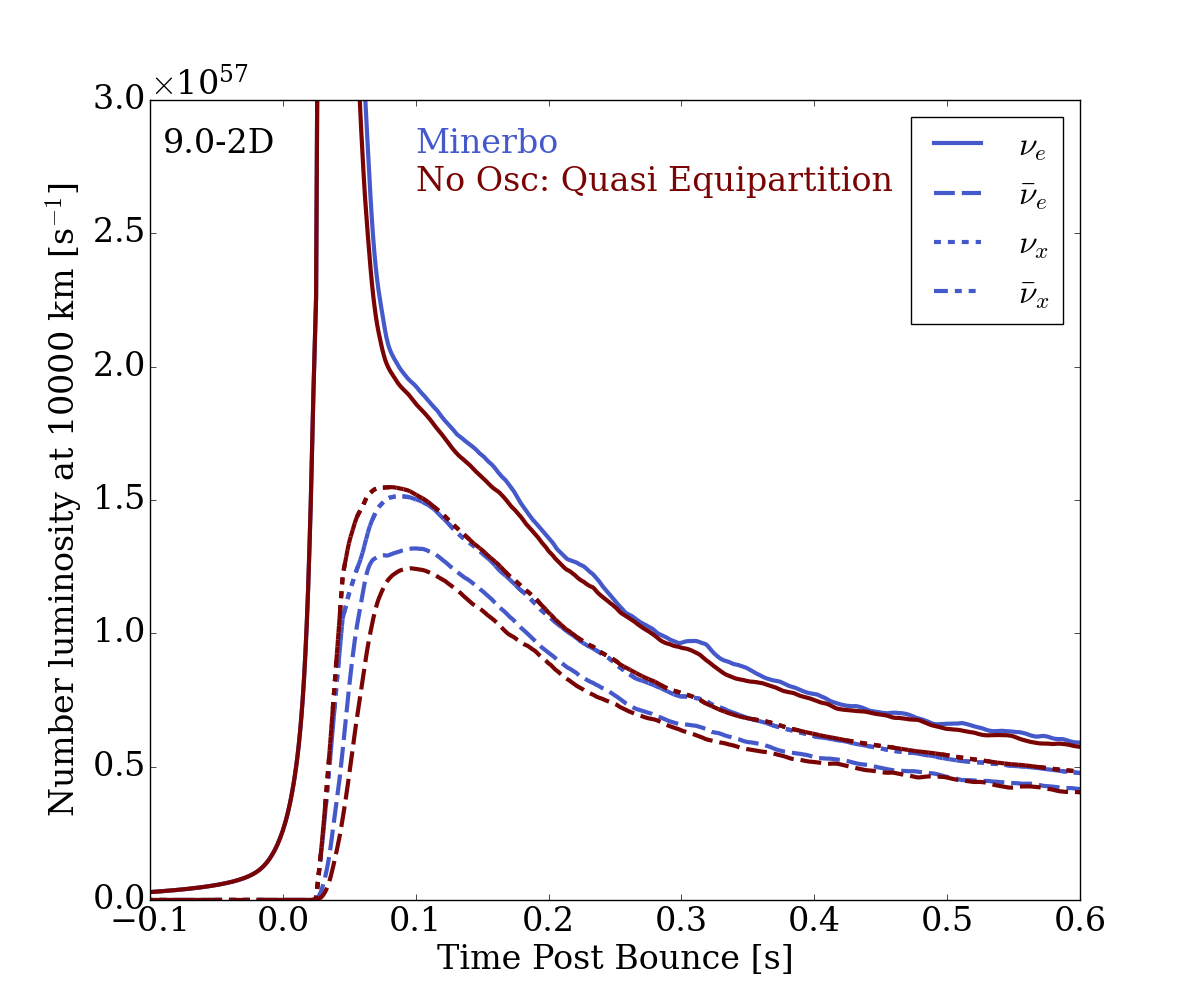}
    \includegraphics[width=0.48\textwidth]{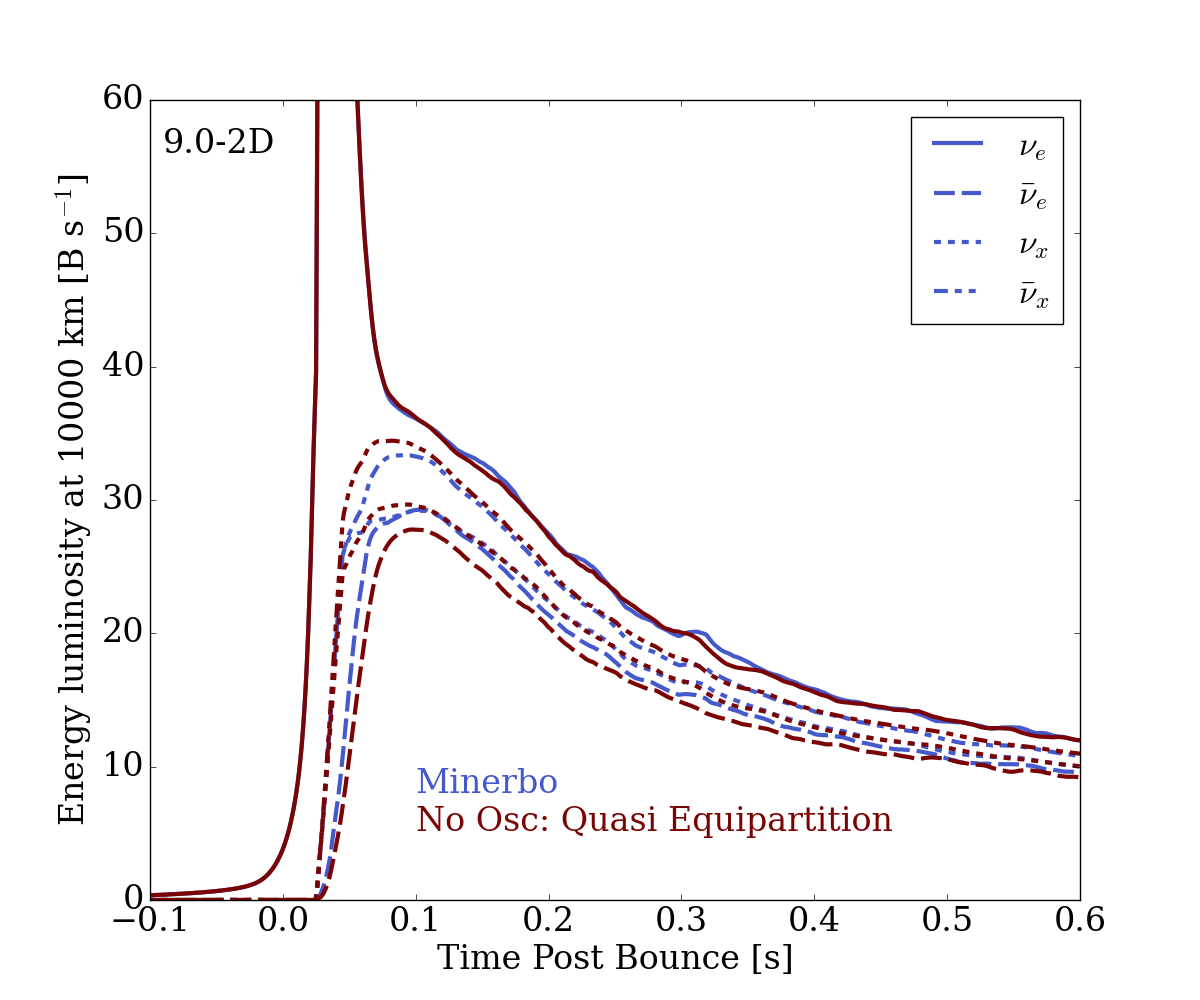}
    \includegraphics[width=0.48\textwidth]{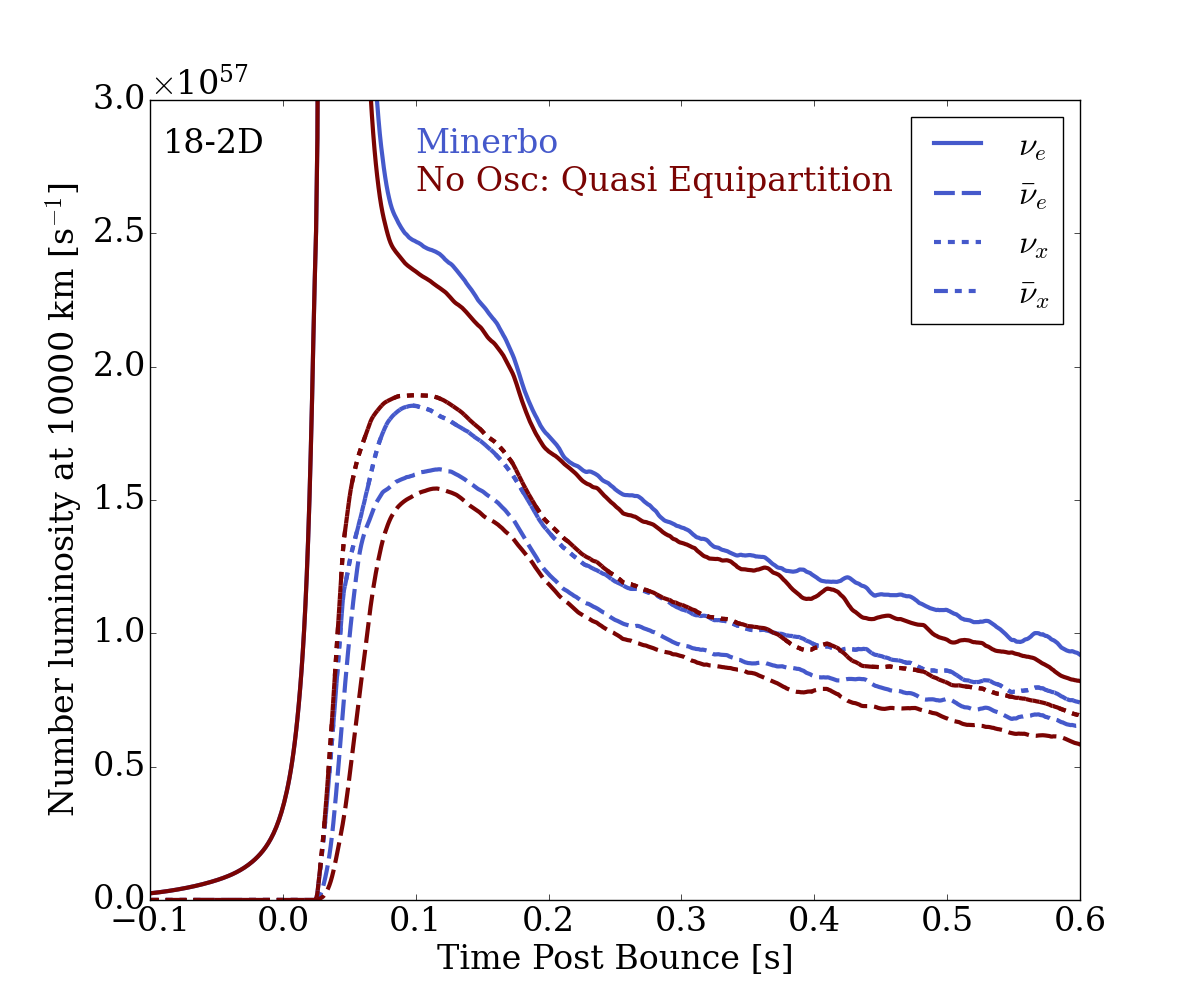}
    \includegraphics[width=0.48\textwidth]{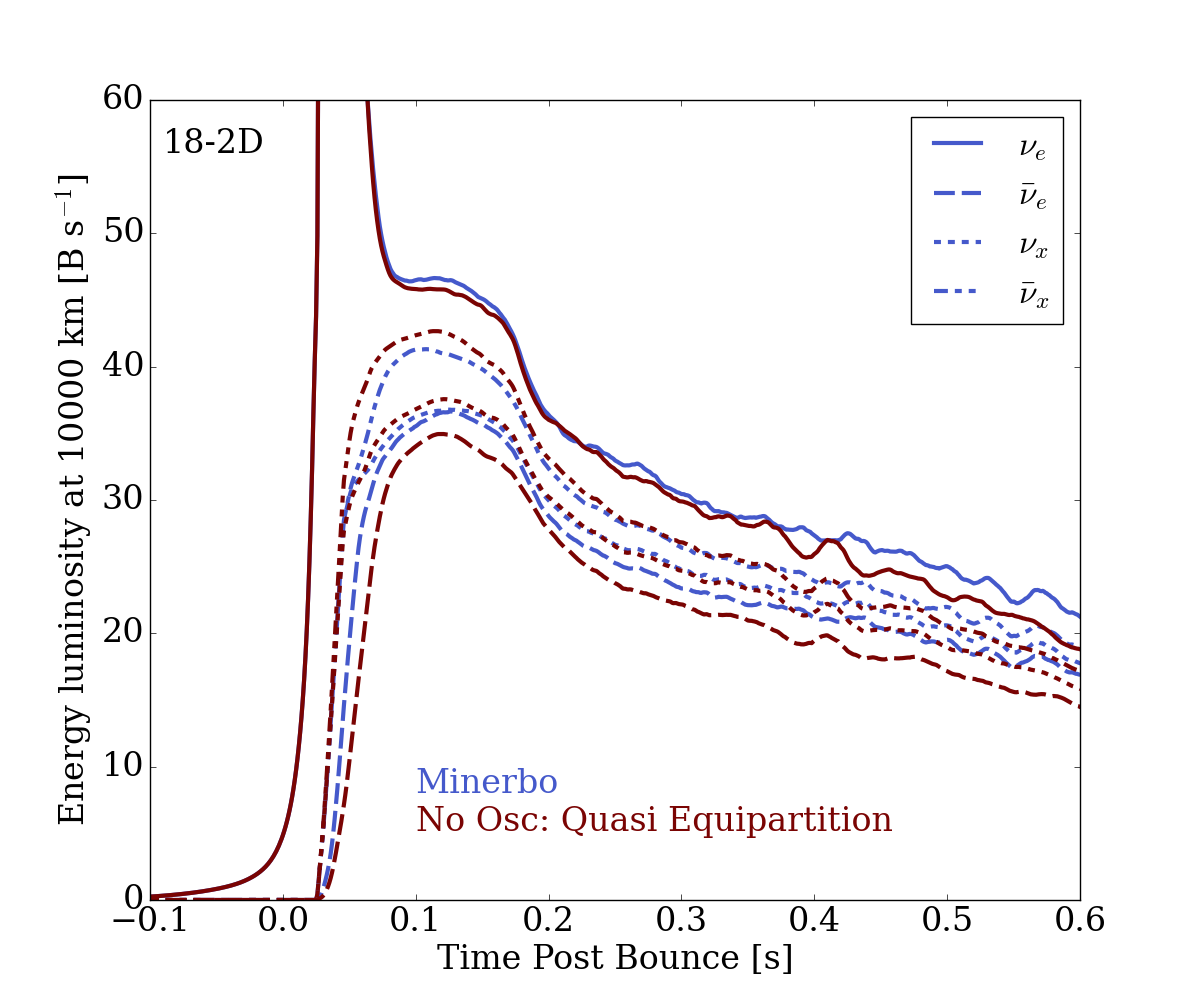}
    \caption{Same as Figure \ref{fig:quasi}, but for the 2D models. Both the neutrino number luminosities and the neutrino energy luminosities are well-approximated by the post-processed quasi-equipartition values from the no oscillation model, especially at late times. The relative errors in 2D are slightly higher than found in the 1D models, but are still less than 10\%. Therefore, the quasi-equipartition approximation is able to capture the major behavior of FFC.
}
    \label{fig:quasi-2d}
\end{figure*}

Figure \ref{fig:quasi-2d} shows the neutrino number and energy luminosities calculated in the 2D FFC models and the post-processed results from the no-oscillation 2D model using the quasi-equipartition formula. Since it can be derived that the quasi-equipartition method conserves $N_{\nu_x}-N_{\bar{\nu}_x}$, the resultant $\nu_x$ and $\bar{\nu}_x$ number luminosities remains identical and, therefore, overlap on the plot. Despite the fact that this phenomenological method cannot capture the number differences between $\nu_x$ and $\bar{\nu}_x$ neutrinos, both the neutrino number luminosities and the neutrino energy luminosities are well-approximated by the post-processed quasi-equipartition values from the no oscillation model, especially at late times. The relative errors in 2D are slightly higher (but are still less than 10\%) compared to the 1D cases, partly due to the stochastic multi-dimensional effects, such as convection and turbulence. But in general, the quasi-equipartition approximation captures the major behavior of FFC, and one can estimate the FFC-altered neutrino properties by post-processing the neutrino signals extracted from no-oscillation CCSN simulations using the quasi-equipartition approximation. This phenomenological method provides a simple way to include the effects of FFC on neutrino signals, without implementing a complex and expensive FFC scheme and re-doing the simulations. Although only tested in CCSN models, this phenomenological method might also work in other FFC-related environments, such as binary neutron star mergers.

\section{Conclusion}
\label{conclusion}
In this paper, we studied the effects of fast-flavor conversion in 1D and 2D core-collapse supernova simulations using a 4-species version of the F{\sc{ornax}} code with the Box3D-BGK method \citep{zaizen2023,richers_box3D.2024,Nagakura2024,wang2025}. We selected two representative progenitors: the 9 M$_\odot$ model from \citet{swbj16} and the 18 M$_\odot$ model from \citet{sukhbold2018}. For each progenitor, we calculated three FFC models using the 4-species scheme with the Minerbo, MEFD, and Nagakura closures, one model using the 3-species scheme with the Minerbo closure, and one model without any flavor conversion. We found that FFC models using different angular reconstruction methods (closures) behave in very similar ways, and that the 4-species scheme results are also similar to those calculated by the 3-species method, except that in the 4-species scheme $\nu_x$ and $\bar{\nu}_x$ have slightly different luminosities. Both 1D and 2D models support this conclusion, and multi-dimensional processes, such as convection and turbulence, don't have a qualitative effect on the character of the FFC effects. 

In all FFC models (1D and 2D), for the low-mass progenitor, the net neutrino heating rates are enhanced by at most $10$\% for a few tens of milliseconds, while the net heating rate changes for the massive progenitor are minor. In 1D, the deviation in shock radius is less than 3 kilometers, which is comparable to the shock radii variation due to closure choices for the classic neutrino transport \citep{wang_burrows2023}. This indicates that the FFC effects are likely to be even weaker. Although the shock radius evolution in 2D is more stochastic due to convection and turbulence, all such models explode at roughly the same times and no significant hydrodynamic effects due to the FFC are found.

The FFC has a stronger impact on the neutrino signals measured at large radii (e.g., 10000 km). Compared to the no oscillation models, about 20\% of the electron and anti-electron neutrino luminosities are converted into the $x$- and anti-$x$-type neutrino luminosities. This conversion fraction decreases as a function of time as the accretion decreases in low progenitor mass models or in successfully exploding massive models. In models that maintain strong accretion at relatively later times, such as non-exploding massive models, the conversion fraction can increase over time. Without the FFC, the $\nu_e$ and $\bar{\nu}_e$ have similar energy luminosities, which are significantly higher than those of $\nu_x$ and $\bar{\nu}_x$ neutrinos. This trend is completely changed by flavor conversion: for models with FFC, the relative differences between energy luminosities of various neutrino types are less than about 10\% after $\sim200$ ms post-bounce, and the luminosity order is in general $L_{\nu_e}>L_{\bar{\nu}_x}>L_{\nu_x}>L_{\bar{\nu}_e}$. The neutrino spectra are also significantly impacted by the FFC. When the FFC is operative, the $\nu_{x}$ and $\bar{\nu}_{x}$ neutrino spectra are softened, while both the $\nu_e$ and $\bar{\nu}_e$ neutrino spectra harden slightly. Neutrinos ($\nu_e$ and $\nu_x$) have similar average energies, while anti-neutrinos ($\bar{\nu}_e$ and $\bar{\nu}_x$) have similar average energies, but at  higher values. 

By plotting the neutrino number density fraction of each neutrino type, we are able to find the location where FFC is happening. We confirm the findings in \citet{wang2025} that the FFC happens mostly in the outer neutrino gain region interior to the shock at early times, while at relative later times, the FFC region interior to the shock shrinks and flavor conversion happens mostly at hundreds to thousands of kilometers radii.

In all our FFC models, we find that the quasi-equipartition condition ($\sqrt{N_{\nu_e}N_{\bar{\nu}_e}}=\sqrt{N_{\nu_x}N_{\bar{\nu}_x}}$) is satisfied after a few hundreds of milliseconds post-bounce. Together with the three conservation constraints of the FFC, this quasi-equipartition condition can be used to derive the neutrino number density and number luminosity after flavor conversion. When applied to the no-oscillation models in a post-processing way, this phenomenological method provides neutrino number luminosities that match the actual FFC models quite well. Together with the assumption that $\langle E^{\rm FFC}_{\nu_e}\rangle=\langle E^{\rm FFC}_{\nu_x}\rangle$ and $\langle E^{\rm FFC}_{\bar{\nu}_e}\rangle=\langle E^{\rm FFC}_{\bar{\nu}_x}\rangle$ after the FFC, the neutrino energy luminosities can be derived as well. Using this quasi-equipartition approximation, one can estimate the FFC altered neutrino properties by post-processing the neutrino signals extracted from no-oscillation CCSN simulations. The relative errors in neutrino number/energy luminosities at late times are less than 2\% for 1D models, and less than 10\% for 2D models. This phenomenological method provides a simple way to include the effects of FFC on neutrino signals without implementing a complex and expensive FFC scheme and re-doing past simulations. However, it is worth noting that the motivation of this phenomenological method is to describe the neutrino properties measured at large radii (e.g., 10000 km), where the flavor conversion has reached an asymptotic state. Since the FFC occurs over hundreds to thousands of kilometers, the neutrino properties at smaller radii will be in an intermediate state between the no-oscillation and the quasi-equipartition states. For studies of neutrino-driven winds or neutrino-related nucleosynthesis, using this phenomenological method might overestimate the effects of the FFC, since such processes are related to neutrino properties at smaller radii. But this method still provides a simple way to describe the FFC-modulated neutrino signals.

We conclude by discussing the caveats and limitations of our calculations. Our comparisons are made here only between 1D and 2D simulations. We haven't looked at how FFC may influence the turbulence field, which can be different in 3D. The Box-like method in \citet{zaizen2023,richers_box3D.2024} is based on QKE simulations with periodic boundary conditions, and the dependence of our results on the boundary conditions needs to be explored \citep{zaizen2023b,cornelius2024}. The Box-like method makes the assumption that there is only one crossing in angular space and that the survival probability function is piecewise constant. Such assumptions need to be further tested and improved. A sensitivity study of the FFC asymptotic/steady state estimation methods is still missing. We ignore heavy leptons in our simulations, while the existence of muons might cause $\nu_x$ and $\bar{\nu}_x$ differences before the FFC and change the outcome. Moreover, three representative closures were used to test the dependence on angular reconstruction methods. However, there is still the possibility that the actual neutrino angular distributions behave fundamentally differently from the reconstructed ones. There is also an inconsistency between the closure-reconstructed neutrino angular distributions and the angular distributions given by the Box3D method after the conversion \citep{richers_box3D.2024}, which needs to be addressed in the future. Therefore, a series of three-dimensional simulations with a multi-angle neutrino transport scheme and a higher-fidelity FFC treatment is desired. 

\section*{Data Availability}  

\adam{The data presented in this paper can be made available upon reasonable request to the corresponding author.}

\section*{Acknowledgments}

TW acknowledges support by the U.~S.\ Department of Energy under grant DE-SC0004658, support by the Gordon and Betty Moore Foundation through Grant GBMF5076, and support through a Simons Foundation grant (622817DK). AB acknowledges former support from the U.~S.\ Department of Energy Office of Science and the Office of Advanced Scientific Computing Research via the Scientific Discovery through Advanced Computing (SciDAC4) program and Grant DE-SC0018297 (subaward 00009650) and former support from the U.~S.\ National Science Foundation (NSF) under Grant AST-1714267. We are happy to acknowledge access to the Frontera cluster (under awards AST20020 and AST21003). This research is part of the Frontera computing project at the Texas Advanced Computing Center \citep{Stanzione2020}. Frontera is made possible by NSF award OAC-1818253. Additionally, a generous award of computer time was provided by the INCITE program, enabling this research to use resources of the Argonne Leadership Computing Facility, a DOE Office of Science User Facility supported under Contract DE-AC02-06CH11357. Finally, the authors acknowledge computational resources provided by the high-performance computer center at Princeton University, which is jointly supported by the Princeton Institute for Computational Science and Engineering (PICSciE) and the Princeton University Office of Information Technology, and our continuing allocation at the National Energy Research Scientific Computing Center (NERSC), which is supported by the Office of Science of the U.~S.\ Department of Energy under contract DE-AC03-76SF00098.

\newpage


\clearpage

\bibliographystyle{aasjournal}
\bibliography{References}

\label{lastpage}
\end{document}